\documentclass[letterpaper, 10 pt, conference]{ieeeconf}  % Comment this line out if you need a4paper

\IEEEoverridecommandlockouts                              % This command is only needed if 
\usepackage{tabularx}
\usepackage{multirow}
\usepackage{todonotes}
\usepackage{cite}
\usepackage{graphicx}
\usepackage{amssymb}
\usepackage{amsmath}
\usepackage{amsfonts}
\usepackage{placeins}

\usepackage{fancyhdr}
\usepackage{makecell}
\usepackage{pifont}

\usepackage{setspace}
\usepackage{amssymb}
\usepackage{tikz}
\usepackage{caption}
\usepackage{subcaption}
\usepackage[stable]{footmisc}
\usepackage{textcomp}

\usepackage{tabularray}

\usepackage{graphicx}
\usepackage{array}

\usepackage{booktabs}
\usepackage{tcolorbox}
\usepackage{tikz}
\usepackage{colortbl}
\usepackage{amsmath}
\usepackage{amssymb}
\usepackage{pifont}
\usepackage{xcolor}
\UseTblrLibrary{booktabs}

\usepackage{textcomp}

\usepackage[linesnumbered,ruled,vlined]{algorithm2e}
\usepackage{graphics} % for pdf, bitmapped graphics files
\usepackage{cite}
\usepackage{url}
\usepackage{hyphenat}
\usepackage{amsmath,systeme}
\usepackage{xcolor}
\usepackage[linesnumbered,ruled]{algorithm2e}
\usepackage{siunitx}
\usepackage{enumerate}
\usepackage{mathtools}

\usepackage[utf8]{inputenc}
\usepackage{ wasysym }
\usepackage{todonotes}

\usepackage{textcomp}

\usepackage{algpseudocode}

\def\floatsep{4.0pt}
\def\textfloatsep{0.0pt}
\def\dblfloatsep{4.0pt}
\def\dbltextfloatsep{4.0pt}

\usepackage{fp}
\FPset{\pb}{0}
\newcommand{\pagebudget}[1]{}

\usepackage{pifont}
\newcommand*\colourcheck[1]{%
  \expandafter\newcommand\csname #1check\endcsname{\textcolor{#1}{\ding{52}}}%
}
\colourcheck{green}
\newcommand{\cmark}{\greencheck}

\def\floatsep{4.0pt}
\def\textfloatsep{0.0pt}
\def\dblfloatsep{4.0pt}
\def\dbltextfloatsep{4.0pt}

\usepackage{fp}
\FPset{\pb}{0}
\renewcommand{\pagebudget}[1]{}

\title{\LARGE \bf A Novel Computer Vision Approach for Assessing Fish Responses to Intrusive Objects in Aquaculture
}

\author{Hanne-Grete Alvheim, Stian Mjelde Jakobsen, Martin Føre and Eleni Kelasidi 
\thanks{Hanne-Grete Alvheim, Stian Mjelde Jakobsen and Martin Føre are with the Department of Engineering Cybernetics, NTNU,  and Eleni Kelasidi is with the Department of Aquaculture, SINTEF Ocean AS.
        [{\tt\small hannegga@ntnu.no}, {\tt\small martin.fore@ntnu.no}, {\tt\small eleni.kelasidi@sintef.no}, \tt\small stian.mjelde20@hotmail.com}
\thanks{This work was supported by the Research Council of Norway through the CHANGE project (NO-313737).}
}
\begin{document}

\maketitle
\thispagestyle{empty}
\pagestyle{empty}

%%%%%%%%%%%%%%%%%%%%%%%%%%%%%%%%%%%%%%%%%%%%%%%%%%%%%%%%%%%%%%%%%%%%%%%%%%%%%%%%
\begin{abstract}

The aquaculture industry needs to address several challenges to secure sustainable seafood production that can serve an increasing global demand. 
One major challenge is to ensure good fish health and acceptable welfare during production since the improvement of fish welfare is of vital importance in current and future production systems. 
In this study, this is addressed by developing and implementing methods to identify fish behaviors in response to intrusive objects both on individual and on a group basis. 
A novel approach for detecting, tracking, and estimating the 3D position of individual fish has thus been developed, and specifically designed to track the caudal fins of farmed fish in industrial sea cages. 
The tracking data was subjected to a novel stereo-vision method adapted to estimate fish positions, velocities, accelerations, and turning and pitch angles. 
Datasets obtained from industrial-scale fish farms were then analyzed to identify the impact of structures of varying shapes, sizes, and colors on fish behavior.\\ 
%A stereo-based vision method was adapted and implemented to estimate fish positions, velocities, accelerations, and turning and pitch angles. 
The method was trained using manually labeled caudal fins, and used YOLOv8 with ByteTrack as an object detector and tracker, SuperGlue for matching detections in the left and right frames, and triangulation to reconstruct the 3D positions of the fish. 
Different image pre-processing and augmentation methods for enhancing object detection accuracy were tested and their performance compared, while RAFT-Stereo was tested for depth estimation purposes. 
The obtained results both validate the method's performance against previous reserach efforts, and demonstrate the novelty and potential of this method in providing more insight into behavioral dynamics in sea-cages.

\end{abstract}
%%%%%%%%%%%%%%%%%%%%%%%%%%%%%%%%%%%%%%%%%%%%%%%%%%%%%%%%%%%%%%%%%%%%%%%%%%%%%%%%
%\input{sections/introduction}
\section{Introduction}
\label{sec:intro}
%A combination of a global increase in the demand for seafood and concerns about the over-exploitation of wild capture fisheries has driven the modern expansion of the aquaculture industry. 
In Norwegian fish farming, Atlantic Salmon (\emph {Salmo Salar} L.) is the most significant species produced both in volume and revenue \cite{salmon_trout}. While the economic yields of salmon farming are large, access to suitable farming sites is scarce, which together with competing claims from other industries and activities in the coastal zone has led to a trend of moving farming operations to more exposed sites \cite{exposed_aquaculture}. In addition, most Norwegian fish farming operations still rely heavily on manual labor, leading to increased personnel risks \cite{Utne2018}. These two factors imply a need for more robust and autonomous solutions as highlighted in the  Precision Fish Farming (PFF) concept \cite{PFF}. The main aim of PFF is to move the industry from an experience-based to a knowledge-based regime, thereby improving accuracy, precision, and repeatability while enabling more reliable decision support and reducing the need for subjective assessments and manual labor. This may, in turn, improve both staff safety and animal welfare at commercial farms. 
PFF concepts have been applied to analyze various aspects of fish farming ranging from feeding optimization \cite{fore2023advanced} to automatic net cleaning and hole detection with robots \cite{fishnetholes}.
%Areas where PFF may be sought utilized can be optimizing fish feeding, which has great economic potential, as fish feeding makes up about half of the cost of salmon farming today, while also being accountable for around 95\% of the carbon emissions \cite{norw_aquaculture_analysis_2019}. Advantageous results can also be applied to reduce biofouling. Automatic net-cleaning robots have been applied succesfully to industrial fish farms in Norway for this scenario. Robots may be utilized to find and inspect holes in nets \cite{fishnetholes}.  
%the need for further expansion to supply a more demanding world population combined with competition with tourism and fisheries has effectively caused a need to apply fish farms to more exposed environments \cite{exposed_aquaculture}.

Unmanned Underwater Vehicles (UUVs) are multi-purpose tools and can be equipped for inspection, maintenance, and repair (IMR) operations and that have several uses in aquaculture, particularly towards solving the aims of PFF \cite{Kelasidi2023}.
UUVs have been utilized in different environments earlier, including the oil- and gas industry, shipping, and oceanography \cite{Kelasidi2023}, suggesting that there are possibilities for utilizing these methods successfully in fish farming environments as well. 
However, industrial fish farms pose challenges as they include the occurrence of large quantities of biomass encapsulated in a small area confined by an elastic and deformable net structure and harsh environmental conditions \cite{Kelasidi2023}. 
Moreover, a UUV will appear as an intrusive object in the cage. 
%The possibility of equipping UUVs with several sensors is a part of moving from a manual to a more autonomous regime \cite{PFF}. This may also provide ways to quantify the behavior of fish. For more exposed fish farming sites, this can be important to ensure well-monitored fish welfare without inducing unnecessary risks to fish farmers. 

Intrusive objects, such as divers, stationary objects, and UUVs, will inevitably have an impact on fish behavior in sea cages. 
Research has been conducted to assess this effect and how it differs between shapes, sizes, and types of objects \cite{kruusma2020, sonar_data}. In \cite{kruusma2020}, experiments studying Atlantic salmon responses when exposed to different moving objects were conducted in an industrial fish farm. 
The objects were a small propeller-driven robot (U-CAT), a remotely operated vehicle ( ROV Argus mini) and a diver, and the metrics compared were the manually estimated distance between the objects and the fish, as well as the tail-beat frequencies of the fish. 
Fish behavior changed when exposed to all three objects and more so when exposed to the diver than to the robots. 
The weakest response was found when exposed to the U-CAT. Tests, where the U-CAT was covered by silver and yellow colors, were also conducted to explore the impact of color, and fish were found to react more toward objects covered in yellow than silver color. 
In a more recent study \cite{sonar_data}, Atlantic salmon responses toward intrusive objects were investigated using deep learning methods applied to sonar data. 
Fish in a commercial fish farm were then systematically exposed to stationary structures of different shapes, sizes, and colors, and sonar measurements were used to estimate the distance the fish kept from the structures. 
8-second scans from the sonar were aggregated, and a deep learning semantic segmentation model was trained to find the fish swimming pattern around the structure, thereafter estimating the smallest distance between structure and fish. 
The fish were found to stay further away from yellow than white structures, and keep a greater distance to larger than smaller structures. The distance was also found to be positively correlated with the weight of the fish. 
In sum, these studies imply that farmed fish are affected by intrusive objects inside sea cages. 
While these responses are predominantly behavioral, they may also entail stress that in turn can lead to reduced welfare.
To enable the design of new robotic solutions and techniques for safe application in fish farming, it is thus necessary to develop new methods to observe and analyze how fish respond to intrusive objects. 
Computer vision methods are likely to be important tools in this regard due to rapid ongoing technological advancements in areas such as AI and processing power.

Computer vision methods, often featuring deep-learning and stereo vision systems have gained traction as monitoring tools in aquaculture over the recent decades \cite{computer_vision_in_fish_farms}. 
%Potential sensors for computer vision applications include monocular and stereo cameras, lidars, and near-infrared cameras. 
%Applications within fish farming may include fish feeding- and sorting, fish counting, biomass estimation, behavioral monitoring, and environmental monitoring. 
Recent advances that have sought to apply 3D distance estimation in sea cages \cite{Aqua3DNet, Nygard2022} may prove particularly relevant for assessing fish responses to intrusive objects.
%which is important in the research phase of putting more robotic solutions in fish farms and observing how it affects the fish. 
For accurate and robust results in cages, such methods must be designed to account for several features including attenuation and absorption of light, fish, feed and feces that occlude images, rapid movements causing motion blurs, and biofouling, all of which may reduce the accuracy and efficiency of 3D position and distance estimation.
In \cite{stereoYOLO} a subset of a video dataset obtained during the trials presented in \cite{sonar_data} was used to demonstrate the effectiveness of a pipeline for tracking multiple fish in 3D by using a stereo camera system. 
The framework consisted of six essential modules: data collection, dataset creation, stereo calibration, object detection, stereo matching, and multiple object tracking (MOT). It was found to be capable of following fish in industrial fish farm settings and estimating both their distance from the camera and their velocity.
The entire fish was detected and tracked, providing a possible way to estimate its quantitative behavior. YOLOv7 was used for object detection purposes, while deepSORT was used to track the fish. 
A similar pipeline for tracking and detecting the distance between tilapia and a given camera system was developed in \cite{koh2023}. 
This system used YOLOv3 for object detection, SORT for object tracking and UDepth for stereo matching and 3D orientation estimation. 
While this system was tested in a small tank of 1m in depth and 2m in diameter rather than in a fish farm, it still showed promise in automating the process of estimating relative fish distance, achieving a mAP50 of 80.63\%. 

In a resembling study, \cite{nygård2022} used YOLOv5, a high-speed multi-object tracker \cite{Bochinski_MOT}, rectification, block matching, sub-pixel refinement, and triangulation to estimate behavioral changes in salmon when exposed to higher levels of CO${_2}$. 
This method focused on the detection and tracking of the eyes and heads of the fish, and synthetic data was used to verify the function of the method. 
Similarly, \cite{Fish_exposed_to_H2S} used a custom-made stereo camera for monitoring the response of Atlantic salmon in recirculating aquaculture systems when exposed to the environmental toxin $H_2S$. 
The smolt exposed to $H_2S$ exhibited a behavioral stress response that was characterized by increased swimming speed, deviating fish patterns, and loss of schooling behavior. 
%In\cite{IoT_digital_twin_fish_farm}, a real-time prototype fish, farming monitoring system, composed of IoT devices, cloud-computing services, and digital twin services integrated with AI was introduced to monitor fish characteristics within a fish farming environment. 
%The digital twin services included fish feeding management, fish metric estimation (weight, size, and count), environmental monitoring, and fish health monitoring. Combining such a system with 3D position estimation techniques may pose a leap forward in the future. 
\begin{figure*}
\includegraphics[scale=0.5]{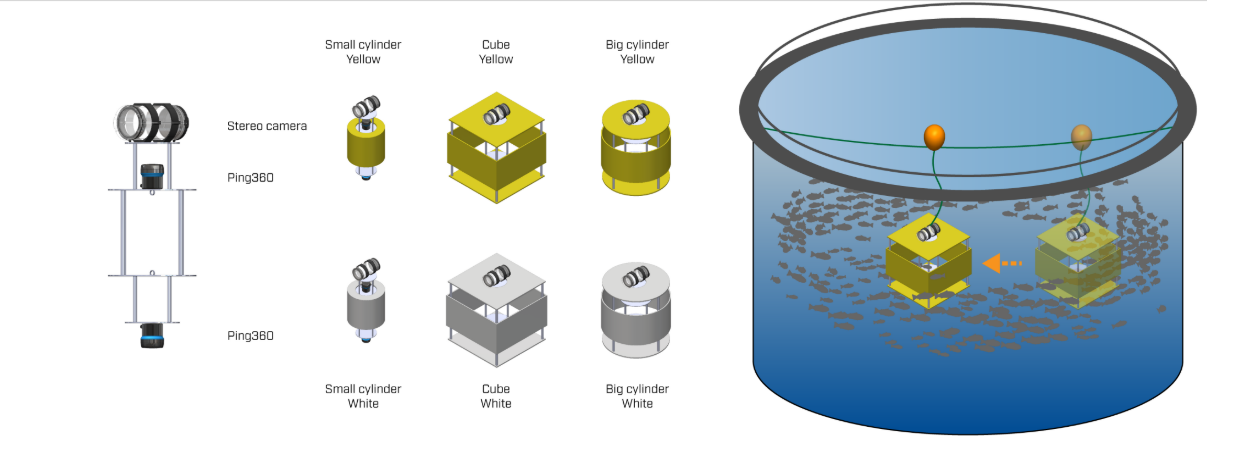}
\caption{Illustration of the experimental setup for data acquisition from industrial scale fish farm of SINTEF ACE}
\label{fig: experimental_setup_visualization}
\end{figure*}

%While all methods introduced above offer something valuable in terms of estimation alternations in fish behaviors when exposed to different objects or stressors, most did not conduct the research in an industrial fish farm. \cite{kruusma2020} estimated the fish behaviour parameters manually, while \cite{sonar_data} used a sonar for data extraction and could therefore not estimate instantaneous behaviour of fish, or behavior other than group behavior. In \cite{stereoYOLO} on the other hand, it was shown that velocity was possible to estimate for single fish. 
The present work sought to develop novel vision-based approach for estimating fish behaviour metrics such as distance, 3D position, velocity, acceleration, turning and pitch angle to study both the local (individual) and the global (group) fish behaviour change in a fish farm.  
The proposed general framework extends the simple pipeline proposed in \cite{stereoYOLO}, but exchanging components therein with new state-of-the-art methods and testing various forms of image pre-processing. 
In particular, in this paper, YOLOv8 \cite{YOLOv8} is applied for object detection, while ByteTrack \cite{ByteTrack} is utilized for multi-object tracking purposes. 
The caudal (tail) fins of the fish were chosen as targets for detection- and tracking purposes. 
SuperGlue \cite{SuperGlue} is used to match features in the left and right frames, while both triangulation and RAFT-Stereo \cite{RAFTstereo} have been tested to consider the depth estimation abilities of the method. 
The method was applied to video recordings in a commercial fish farm \cite{SINTEF_ACE} and was proven able to accurately estimate fish behaviour metrics in industrial fish farming environments.

The best performance was achieved when the object detector was trained on data with labeled caudal fins for fish in both the foreground and background of the images, using triangulation instead of RAFT-Stereo (termed the \emph{Initial augmented} method), and when images were pre-processed using morphological area opening and wavelet transforms (termed \emph{MO-WT augmented}). 
These two methods were then applied to video data obtained from field trials aimed at exploring fish responses toward stationary intrusive objects of different sizes, shapes, and colors to gauge the relative distance between fish and the objects. 
Statistical analyzes implied that the fish kept shorter distances to white and small objects and longer distances to yellow and large objects. These outcomes harmonize with results from previous studies with similar aims \cite{sonar_data}, underlining the importance of considering such fish-object dynamics in the future design of underwater vehicles and navigation methods for aquaculture. 
The methods developed in this study are highly relevant for further research on stereo vision in fish farming. % and are expected to contribute to increased welfare in industrial-scale fish farms.

\section{Experimental setup and Data Acquisition}
\label{sec:exp}

Data was collected at the SINTEF ACE facility \cite{SINTEF_ACE} Korsneset on the 6th of September, 2022. 
Average fish weight was 1084 g at the time of the experiment. 
A structure equipped with a custom-made stereo camera and two Ping360 sonars was placed around 5 m from the edge of the cage, at a depth of 8 m, and kept at the desired location being suspended from a rope attached to the floating collar and a buoy. 
%The ropes spanned the cage and were tied to the buoy, and the structure hung freely from one rope. 
%Between any disturbance of the structure, a waiting period of at least 10 minutes was taken to reduce its potential impact on fish behavior. 
The structure was coated in three different shell structures: a small cylinder with dimensions Ø30 x 30 cm, a large cylinder with dimensions Ø60 x 60 cm, and a cube with dimensions 60 x 60 x 60 cm. The structures were coated with white and yellow sheets. This resulted in a total of six different combinations as visualized in Figure \ref{fig: experimental_setup_visualization}.

Six 12-minute videos were recorded for each structure-color combination, moving the structure laterally for approximately 25 s between repetitions to reset the situation. 
After each movement of the structure, the system was left inactive for 10 minutes before starting the recording to ensure that the transient response of the fish caused by the motion had subsided. 
To further reduce the risk that transients would impact the results, the the first and last minutes of the 12-min videos were not used in the analyses. 
All replicates were used in the analyses except 4 replicates of white cube (cleaner fish obstructed the view, and 1 replicate for small white cylinder that was used to train the object detection model.
% For all structures apart from the white cubical structure and the small white cylindrical structure, all six replicates were utilized for testing. 
% In the case of the white cube, a cleaning fish obstructed the view in four of the videos, deeming the results too unreliable for further analysis. 
% One video from the small white cylinder was used for training the object detection model and, therefore, not utilized in the analysis (Table \ref{table:experiment_setup}). 

The stereo camera used in the experiments is custom-made and built using two Lucid TRI032S-CC GigE vision cameras. 
The two cameras were set 42 mm apart inside a 3D-printed bracket in a BlueRobotics 4" watertight housing with a flat glass window. 
%Further specifications can be found in \cite{stereoYOLO}.  in parallel
Prior to the experiments, stereo camera calibration was conducted using underwater images of a chessboard plane with squares of known size. Further details about the system specifications and calibration process can be found in \cite{stereoYOLO}. 
Stereo mapping from the 3D plane to the 2D plane, along with camera specifications \cite{disparity_mapping} can then be used to calculate the depth and 3D position of objects captured by the camera. 

\begin{table}
\centering

\resizebox{\textwidth/2}{!}
{\begin{tblr}{
  vline{3} = {-}{},
  hline{1,2,4,6,8} = {-}{},
}
 \textbf{Shape} &  \textbf{Color} & \textbf{Location}  & \textbf{Date} & \textbf{Replicates} \\
 Big Cylinder & White & Korsneset & 5-Sep 2022 & 6\\
 & Yellow & Korsneset & 6-Sep 2022 & 6\\
  Cube & White & Korsneset & 7-Sep 2022 & 2 \\
 & Yellow  & Korsneset & 7-Sep 2022 & 6 \\
 Small Cylinder & White & Korsneset & 8-Sep 2022 & 5\\
 & Yellow  & Korsneset & 8-Sep 2022 & 6

\end{tblr}}
\caption{Comparison of methods in terms of average distance}

\label{table:experiment_setup}
\end{table}
\begin{figure*}
    \centering
    \includegraphics[scale=0.62, trim={0cm 1.2cm 0cm 2.5cm},clip]{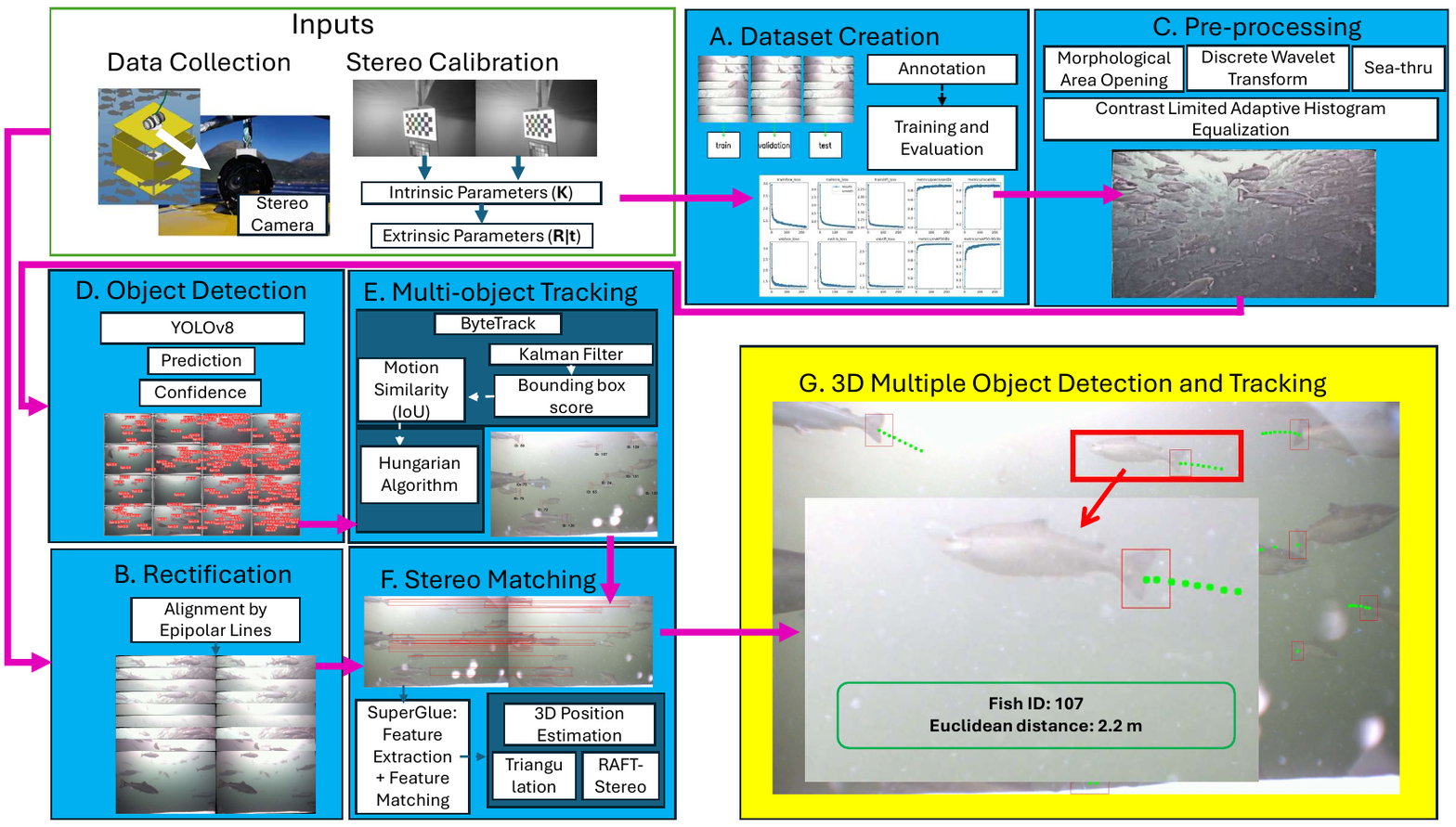}
    \caption{A block diagram depicting the entire process from capturing data to tracking and estimating fish positions. Inputs: data collection and stereo calibration, A. dataset creation, B. rectification, C. pre-processing, D. object detection, E. multi-object tracking, F. stereo matching, and G. 3D multiple object tracking and detection. The core object detection algorithm used is YOLOv8 and ByteTrack is used for MOT tracking. The pink lines represent data transferred from one module to another.}
    \label{fig:complete_workflow}
\end{figure*}

\section{Pipeline for stereo video processing}
\label{sec:app}

%In this study, a framework for detecting and estimating the depth of fish is proposed, influenced from the framework in \cite{stereoYOLO}. 
Figure \ref{fig:complete_workflow} provides an overview of the overall structure of the pipeline used to acquire the 3D distances from the structure. 
The calibration procedure (see \cite{stereoYOLO} for details) and the collected dataset are inputs to the system. 
Thereafter, the system follows several steps: dataset creation and annotation (A), rectification (B), pre-processing (C), object detection by YOLOv8 (D), multi object tracking by ByteTrack (E), stereo matching with SuperGlue (F) and finally the last step, 3D multiple object detection- and tracking (G). 
The steps are tailored to work in underwater environments subject to heavy visual occlusion due to fish and particles. Each step in the proposed framework is explained further in detail in subsequent subsections. 

\subsection{Dataset Creation}
The dataset for training, validation, and testing was based on individual frames from one of the videos recorded while the fish were exposed to the small yellow cylinder. 
Since consecutive frames in a video stream with 24 FPS will look very similar, training the model based on a sequence of consecutive images could lead to overfitting. 
To avoid this, every 50th frame was extracted from the video and included in the dataset. 
The dataset was built using only the left frame of the stereo camera system, and resulted in a total of 686 images that were split into separate training (approximately 60\%), validation (20\%) and test (20\%) sets. Python and OpenCV were used to create the dataset, and annotation was done using CVAT to label the caudal fins of the fish in the foreground of the video (i.e., only those fish closest to the camera). 
This dataset is referred to as the \textit{Foreground caudal fins dataset}. 

Since the accuracy of AI methods for automatic detection depends on the quality of the training data, it is relevant to investigate if more extensive and elaborate labeling may lead to improved performance. 
A second dataset consisting of annotated caudal fins of fish in both the foreground and the background was therefore also created based on another video taken at a different cage. 
For this dataset, one frame was extracted from the video every five seconds for the training set, resulting in a total of 429 images. 
The validation and test sets from the first dataset were used for validation and testing of the methods trained using the second dataset. This dataset is referred to as \textit{Complete caudal fins dataset}.
Labelling of the caudal fins in both background and foreground was done using MakeSense. 

\subsection{Rectification}

\begin{figure}
    \centering % <-- added
    
\begin{subfigure}[t]{0.50\textwidth}
\centering
  \includegraphics[scale=0.06]{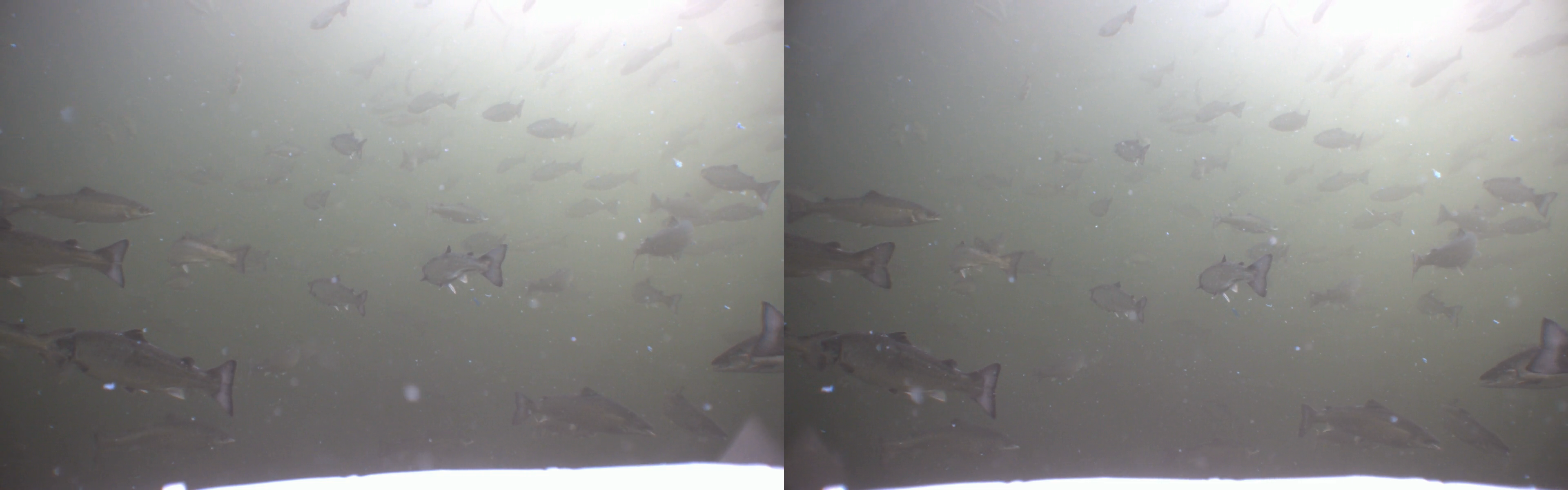}
  \caption{}
  \label{fig:non_rect_img}
\end{subfigure}\hfil % <-- added
\begin{subfigure}[t]{0.50\textwidth}
\centering
  \includegraphics[scale=0.06]{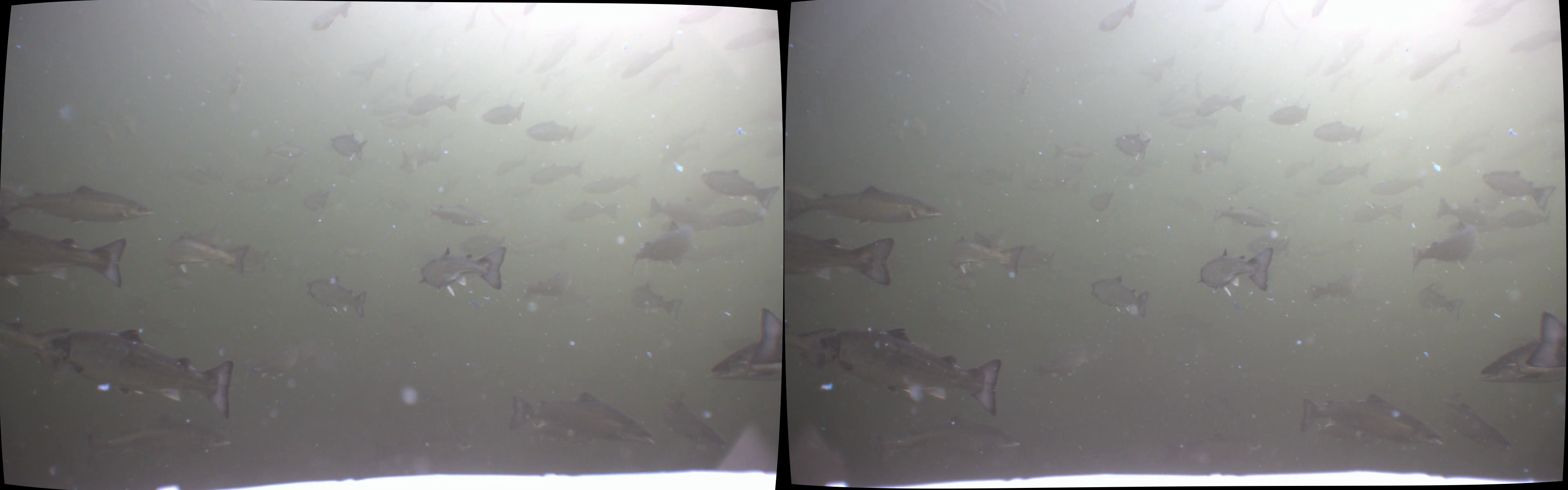}
  \caption{}
  \label{fig:rectified_img}
\end{subfigure}\hfil % <-- added
\caption{A non-rectified frame (a) compared to a rectified frame (b)}
\label{fig:rect_ex}
\end{figure}

After the calibration procedure has been conducted for the stereo camera system, the intrinsic parameters of each camera and the relative position between the left and right camera is captured. 
The intrinsic and extrinsic parameters are then used in the rectification procedure \cite[p.~302]{Image_geometry_stereo}, which translates and rotates pairs of images, such that the epipolar lines run collinearly and parallell to a chosen image axis (typically the horizontal axis). 
The rectification process restricts the searching area for corresponding points, and is therefore used in this approach. 
Rectification was done using the remap function from OpenCV, where two rectification maps providing pixel-wise mappings are provided as inputs. A visualization of the rectification procedure for a frame is shown in Figure \ref{fig:rect_ex}. 
The minimal baseline distance between the two cameras results in very similar views, causing only minor changes during the rectification process. 

\subsection{Pre-processing}
\begin{figure*}[h!]
    \centering % <-- added
    
\begin{subfigure}[t]{0.50\textwidth}
\centering
  \includegraphics[scale=0.1]{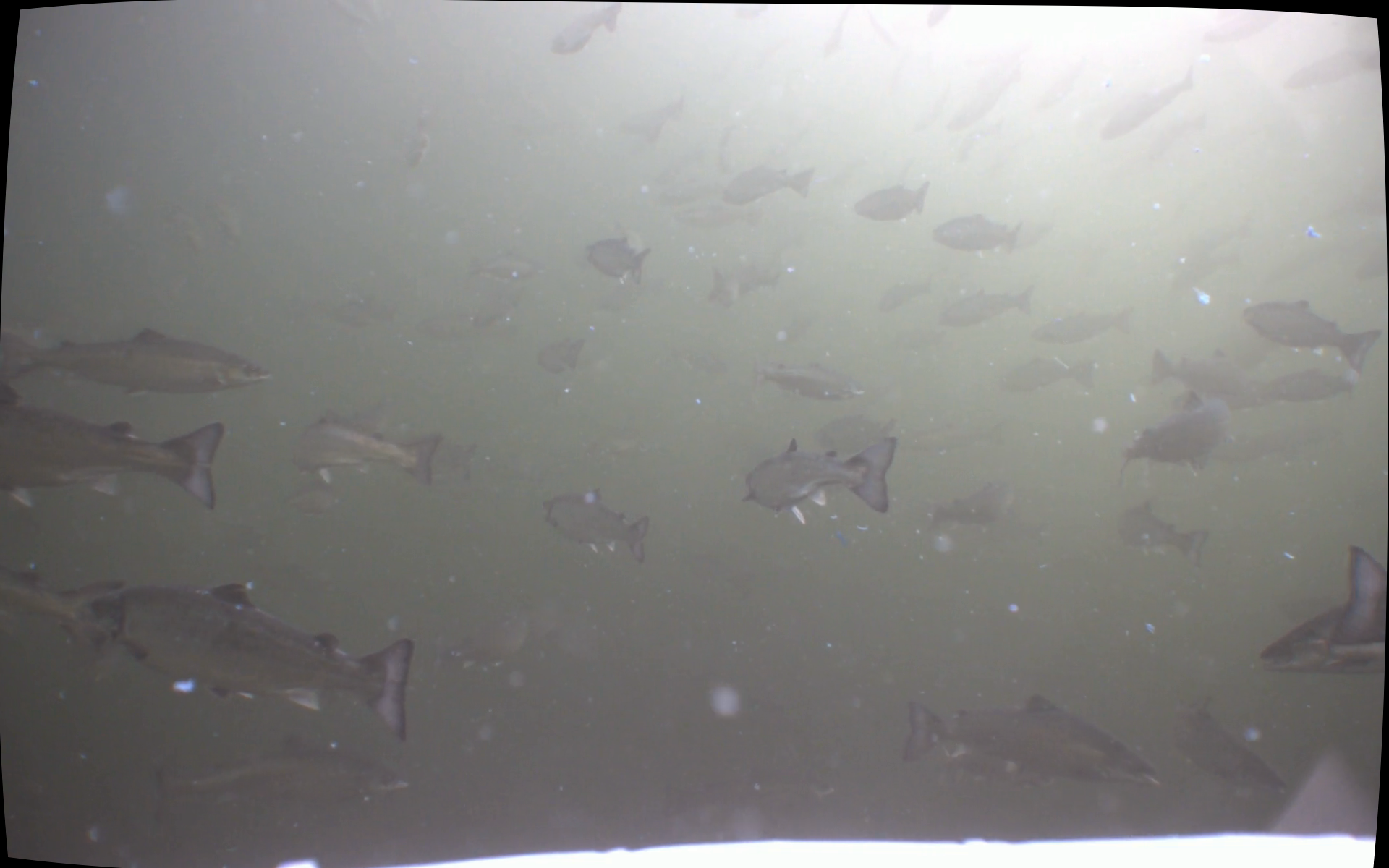}
  \caption{Initial}
  \label{fig:initial_img}
\end{subfigure}\hfil % <-- added
\begin{subfigure}[t]{0.50\textwidth}
\centering
  \includegraphics[scale=0.1]{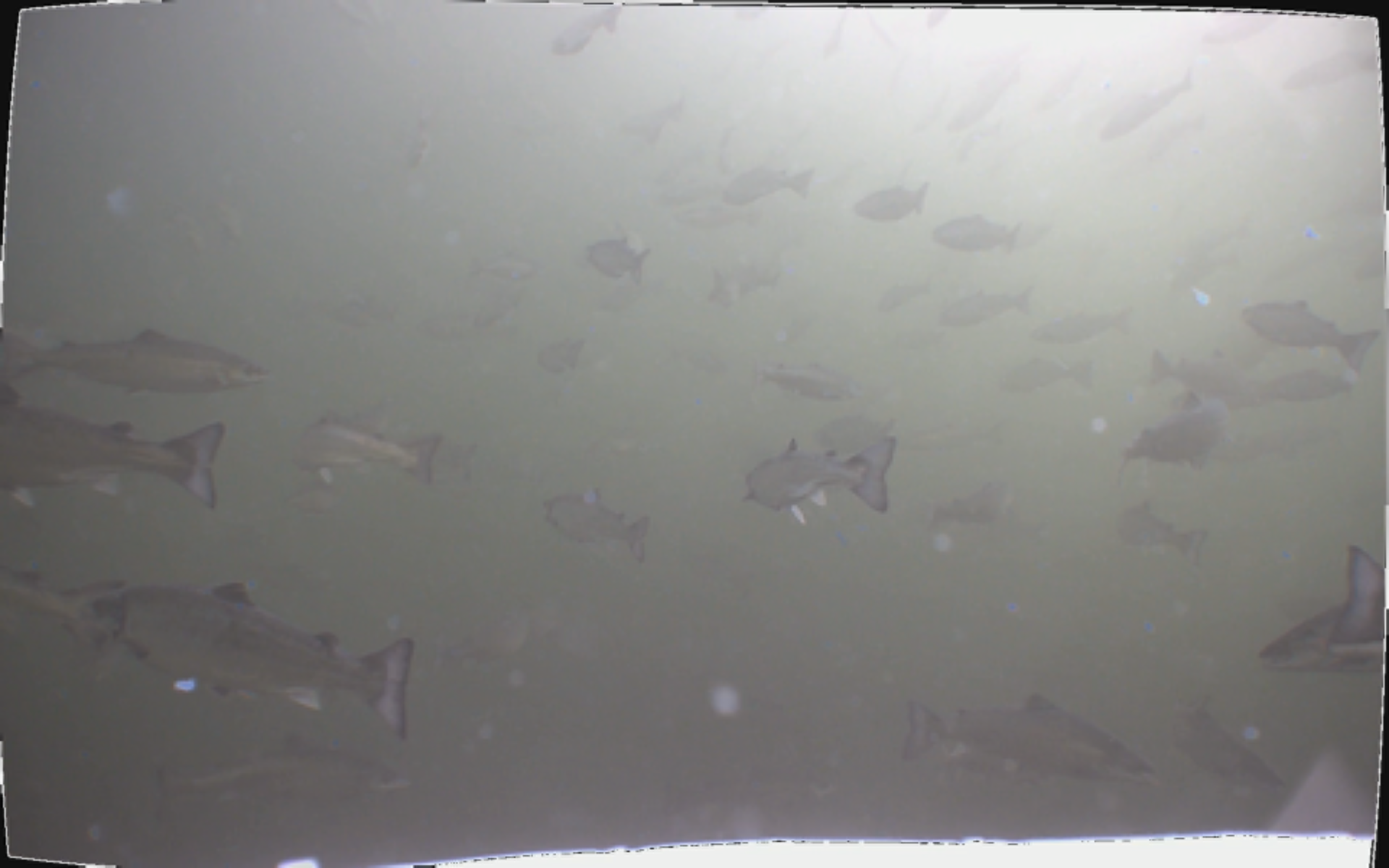}
  \caption{MO-WT}
  \label{fig:mo_wt}
\end{subfigure}\hfil % <-- added
\begin{subfigure}[t]{0.50\textwidth}
\centering
  \includegraphics[scale=0.1]{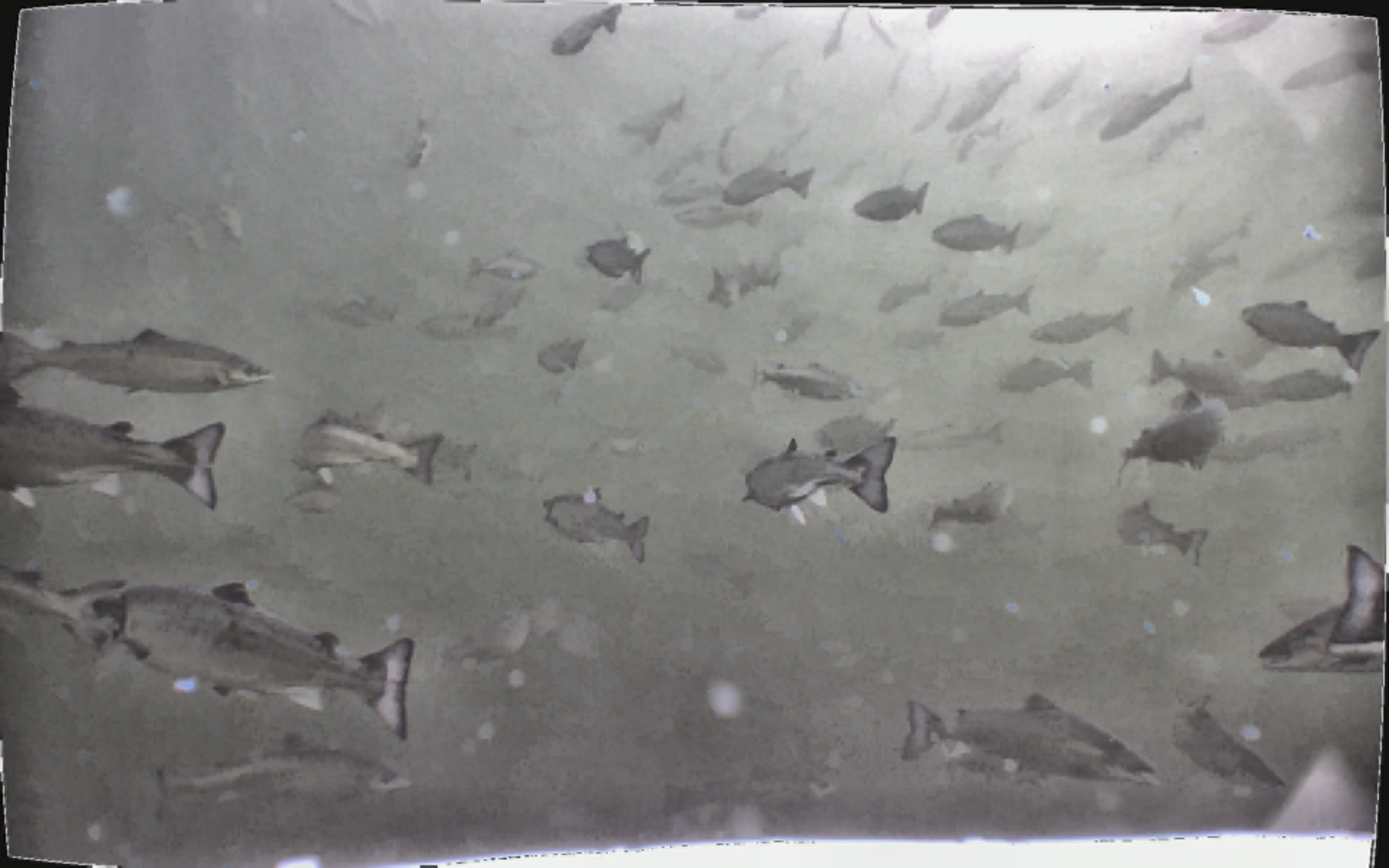}
  \caption{MO-WT and CLAHE}
  \label{fig:mo_wt_clahe}
\end{subfigure}\hfil % <-- added
\begin{subfigure}[t]{0.50\textwidth}
\centering
  \includegraphics[scale=0.07]{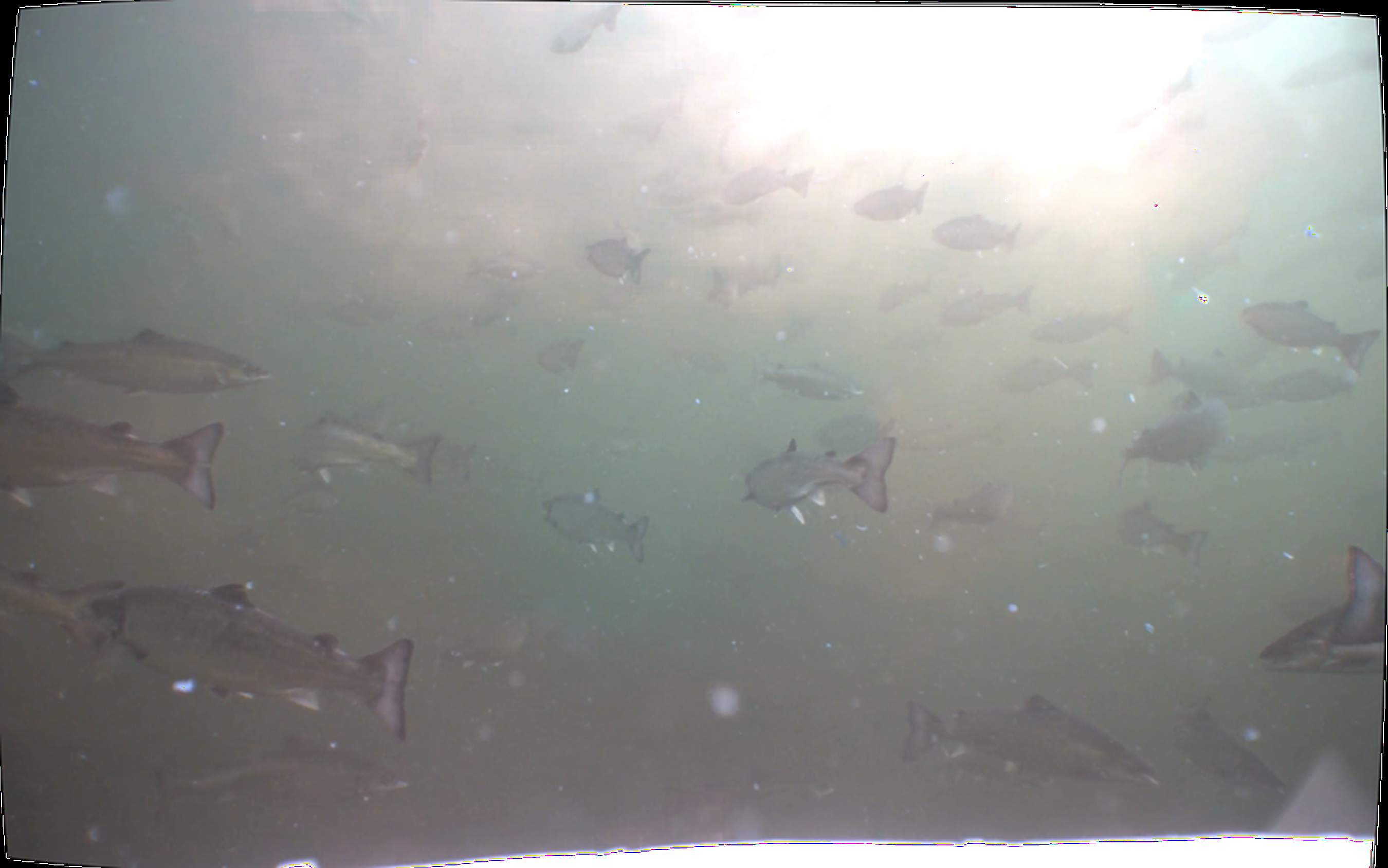}
  \caption{Sea-thru}
  \label{fig:sea-thru}
\end{subfigure}\hfil % <-- added
\caption{A single left frame altered by different pre-processing approaches.}
\label{fig:pre-processing_ex}
\end{figure*}

Visibility in industrial fish farms may be influenced by severe occlusion, noise, and large particles due to the presence of farmed fish, cleaner fish, structural components in the farm and influxes from the ambient environment. 
This may further complicate the ability to reliably estimate 3D positions of fish. 
Fewer fish may be detected, and position estimates may also become less reliable. 
Image improvement techniques were therefore tested as pre-processing methods to counteract this. 
The methods that were tested were discrete wavelet transformation (DWT) \cite{WT}, morphological area opening (MO) \cite{Morph_area_opening}, contrast limited adaptive histogram equalization (CLAHE) \cite{CLAHEOriginal} and Sea-thru \cite{Sea_thru}. 
%All these methods were assumed to contribute to enhancing the underwater images. 
The effect of these different image improvement techniques on a single frame is illustrated in Figure \ref{fig:pre-processing_ex}, where "Initial" refers to the unmodified raw image, "MO-WT" the image after the application of MO and DWT, "MO-WT and CLAHE" the image resulting from the application of CLAHE after MO and DWT, and "Sea-thru" the image produced by applying Sea-thru to the raw image.  

\subsubsection{DWT}
DWT reduces noise in an image without impacting the sharpness around the edges. An image is decomposed into several wavelet coefficients, which are used to represent the details of the image in various ways concerning scale and resolution. Similar to a filtering mechanism, the image is subjected to one or more pass-band, high-pass and low-pass filters to find the wavelet coefficients. 
The value of the different wavelet coefficients represents how well a wavelet approximates a portion of the image in question, with large wavelet coefficients implying good approximation and low coefficients meaning that the wavelet does not represent the image well and is either noise or a non-informative part of the image. 
Removing wavelets with low coefficients from the image will therefore reduce noise in the image without reducing image information details. The discrete wavelet transform is implemented with a library for Python known as PyWavelets, using Daucebies wavelets with a decomposition level of 2. 
%The value of the different wavelet coefficients represents how well a wavelet approximates a portion of the signal in question, where a large wavelet coefficient would imply a good approximation. A low coefficient, on the other hand, means that the wavelet does not represent the signal well and is either noise or a non-informative part of the signal. By removing these wavelets from the signal, noise is therefore reduced in the image without reducing image information details.

\subsubsection{MO}
MO can be used to remove large particles from the image. This transform employs a structuring element of a specified size to represent a neighborhood. It uses a threshold to identify and remove lighter blobs that exceed a certain size in the image. MO is implemented by using the AreaOpening function from the Python wrapper of the DIPlib C++ library \cite{diplib}.

\subsubsection{CLAHE}
CLAHE is a version of histogram equalization, aimed at improving the overall contrasts of images. The approach creates a histogram describing the luminance of an image, and uses this information to make all the different intensities within the image equally probable. 
This improves contrast in an image, but has a drawback as it may suppress local edges and features in favor of more dominant features in the image. 
CLAHE processes the image locally with a window size and introduces a threshold on the amplification value of the histogram called the clip limit, which limits the possible amplification of noise caused by histogram equalization. CLAHE is implemented with OpenCV, and uses the LAB color domain to estimate luminance.

\subsubsection{Sea-thru}
Sea-thru is an algorithm especially created for underwater color correction, denoising and dehazing, with an overall goal of removing the effects of the water from the image. 
Sea-thru is a model-based approach, using a revised version of the commonly used atmospheric image formation model. By making certain assumptions regarding depth between the stereo camera system and objects in the image, parameters can be placed into the modelled underwater world to color-correct, dehaze and denoise the image. 
An open-source version of Sea-thru \cite{sea_thru_implementation} was used for these experiments, the depth map being created using RAFT-Stereo and its real-time architecture. 

\subsection{Object detection}
\begin{figure}[h!]
    \centering % <-- added
    
%\begin{subfigure}[t]{0.50\textwidth}
%\centering
  \includegraphics[scale=0.1]{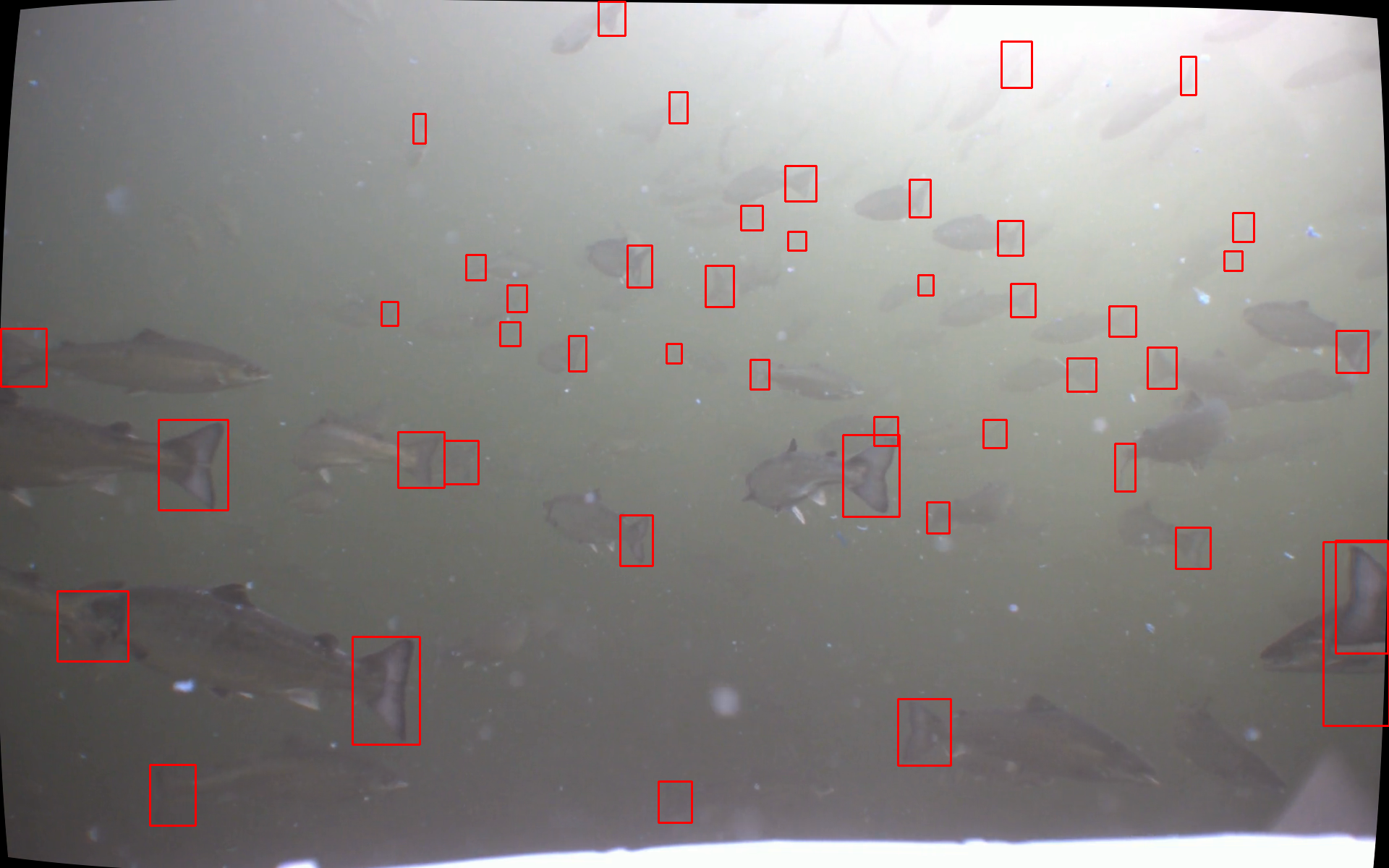}
%  \caption{Initial}
%  \label{fig:initial_img}
%\end{subfigure}\hfil % <-- added
%\begin{subfigure}[t]{0.50\textwidth}
%\centering
%  \includegraphics[scale=0.06]{Method_figures/obj_det/object_detections_mo_wt.png}
%  \caption{MO-WT}
%  \label{fig:mo_wt}
%\end{subfigure}\hfil % <-- added
%\begin{subfigure}[t]{0.50\textwidth}
%\centering
%  \includegraphics[scale=0.1]{Method_figures/obj_det/object_detections_clahe_mo_wt.png}
 % \caption{MO-WT and CLAHE}
  %\label{fig:mo_wt_clahe}
%\end{subfigure}\hfil % <-- added
%\begin{subfigure}[t]{0.50\textwidth}
%\centering
%  \includegraphics[scale=0.045]{Method_figures/obj_det/object_detections_sea_thru.png}
%  \caption{Sea-thru}
%  \label{fig:sea-thru}
%\end{subfigure}\hfil % <-- added
\caption{Object detection on a single left frame}
\label{fig:objectdetection_ex}
\end{figure}

YOLOv8 is a state-of-the-art real-time object detector known for its high accuracy and was used to detect bounding boxes around caudal fins. 
The models in the 'You Only Look Once' series are known for their real-time capabilities, and as the models have been incrementally updated they have lately become known for their accuracy as well. 
Like its predecessors, YOLOv8 uses a single convolutional neural network (CNN) to simultaneously delegate class probabilities and the location of bounding boxes.
The bounding boxes are specified by the center coordinate (x, y) and the width and height of the box (w, h), while a scalar value presents the confidence score. YOLOv8 has an anchor-free architecture, enabling it to detect different object sizes better than its predecessors. Moreover, it uses a modified version of CSPDarknet53 as the backbone. 
For these experiments, the detector only detects tail fins in the left frame of the stereo vision system using the smallest model, YOLOv8n, with default augmentation (hereafter termed \textit{augmented}) which increases the image width to 1024 pixels, applies vertical image flipping with a probability of 0.4 and mixup with a probability of 0.3. 
%Different methods of augmentations are also tested apart from the default augmentation. 
%This augmentation of the default settings is hereafter termed \textit{augmented}. 
Training, validating and testing the different object detectors was done using the Ultralytics YOLO module. The same module was used for predicting the placement and number of fishtails in the complete framework. Figure \ref{fig:objectdetection_ex} depicts the object detection procedure on a single frame. 

\subsection{Multi-object Tracking}
\begin{figure}[h!]
    \centering % <-- added
    
%\begin{subfigure}[t]{0.50\textwidth}
%\centering
  \includegraphics[scale=0.1]{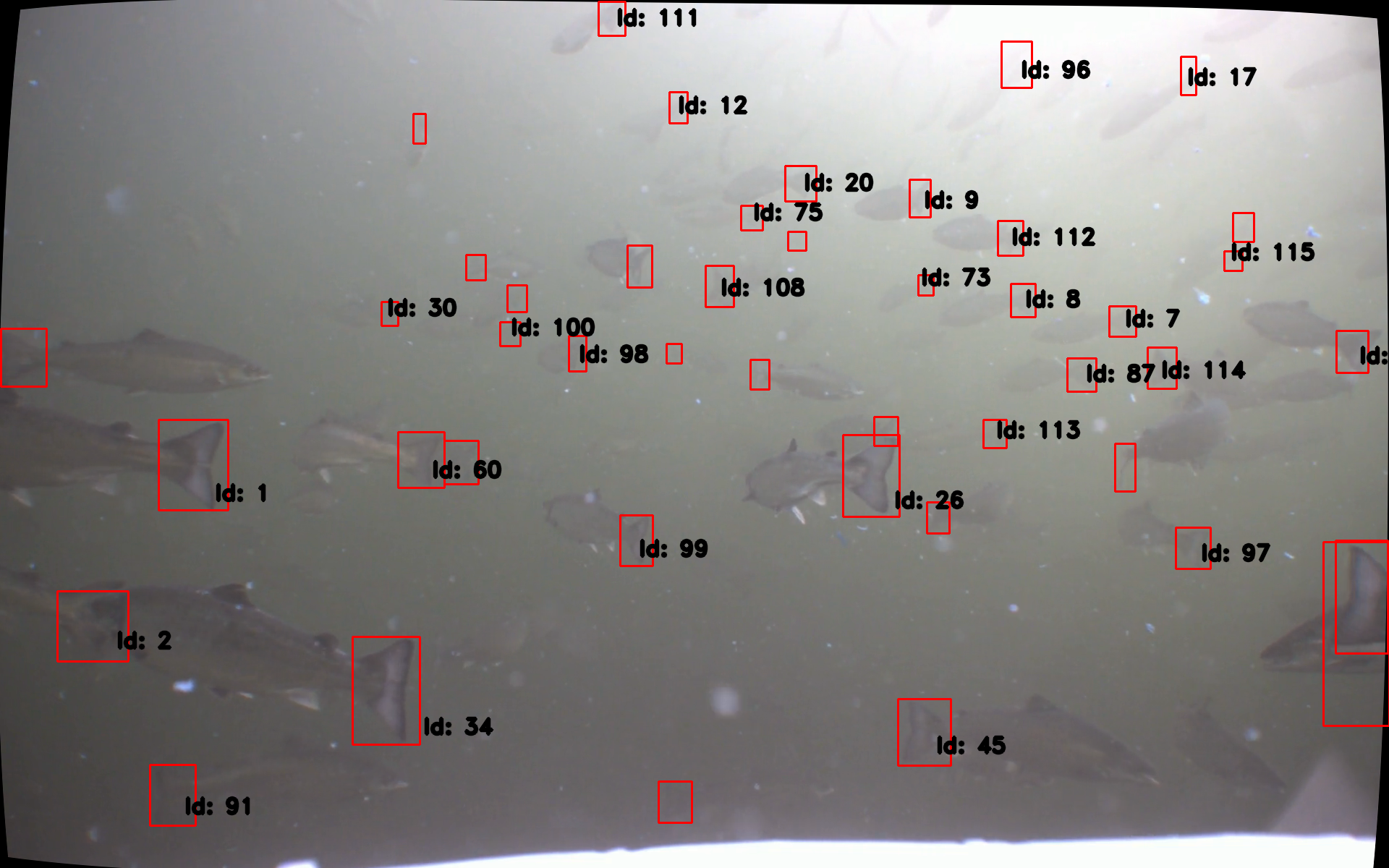}
 % \caption{Initial}
 % \label{fig:initial_img}
%\end{subfigure}\hfil % <-- added
%\begin{subfigure}[t]{0.50\textwidth}
%\centering
%  \includegraphics[scale=0.06]{Method_figures/MOT/object_detections_tracks_mo_wt.png}
%  \caption{MO-WT}
%  \label{fig:mo_wt}
%\end{subfigure}\hfil % <-- added
%\begin{subfigure}[t]{0.50\textwidth}
%\centering
%  \includegraphics[scale=0.1]%{Method_figures/MOT/object_detections_tracks_clahe_mo_wt.png}
  %\caption{MO-WT and CLAHE}
  %\label{fig:mo_wt_clahe}
%\end{subfigure}\hfil % <-- added
\caption{MOT on a single left frame}
\label{fig:multiobjectdetection_ex}
\end{figure}

Ultralytics also includes a multi-object tracking (MOT) module. 
ByteTrack is one of the MOT algorithms offered by this module, and was chosen for this study due to its real-time abilities and generally high accuracy. 
%This was necessary to capture the distances to the fish and derived features such as swimming velocity, which may only be measured when fish move actively within the camera frame and are assigned unique identifiers for accurate tracking. 
Unlike classical MOT methods, which only consider detection boxes with a score above a certain threshold, ByteTrack considers all detection boxes. It starts by associating tracks with detection boxes with high confidence based on the IoU score or similarity score. As this process is accomplished, the remaining tracks that are not yet associated with a detection box are associated with low-confidence detections based on the IoU score. The Hungarian algorithm solves both associations. 
This process could be highly beneficial in fish farming environments, where rapid movements and occlusions may cause blurred or occluded detected fishtails, possibly reducing similarity scores or IoU scores. Figure \ref{fig:multiobjectdetection_ex} visualizes the MOT approach using ByteTrack on a single frame. 

\subsection{Stereo Matching}
To capture the center of the caudal fin in both frames, the deep learning method Superglue was used for stereo-matching purposes by establishing point correspondences between the two frames captured by the stereo vision system. SuperGlue relies on a CNN termed SuperPoint \cite{Superpoint}, which detects points of interest in images and their associated visual feature descriptors. Moreover, it combines graph-based methods with optimization in a problem known as the optimal transport problem, utilizing transformer theory \cite{Transformers}, and an improved version of the Hungarian algorithm \cite{Hungarian_algorithm} to gather corresponding points and information about these. Associated points in each frame are combined with the center points from the bounding box to estimate the center of the tail of the fish in the right frame. 
When there are no point correspondences, detections are discarded. 

%\subsubsection{Triangulation}
Depth estimation is done by testing two different approaches, the first of which uses triangulation to estimate the 3D position of the fish. 
This involves using the disparity information between matching corresponding points found by the SuperGlue approach in combination with calculated intrinsic parameters to estimate the 3D positions of the point. 
%\subsubsection{RAFT-Stereo}
The second approach utilizes RAFT-Stereo to estimate the relative distance between the fish and the camera in real-time. 
RAFT-Stereo is a deep learning architecture based on the optical flow method RAFT \cite{RAFT}, with a primary objective to combine depth estimation with optical flow principles to estimate a displacement map for all pixel values of a rectified image in a stereo vision system. 
While RAFT-Stereo has been shown to generate robust and reliable results in the most common benchmarks for stereo vision tasks \cite{Kitti,Middlebury}, it has not been trained in underwater scenarios. This may pose challenges due to differences in color- and lightning variations along with distortion effects. 

\begin{figure*}
    \centering % <-- added
    
%\begin{subfigure}[t]{\textwidth}
%\centering
  \includegraphics[scale=0.126]{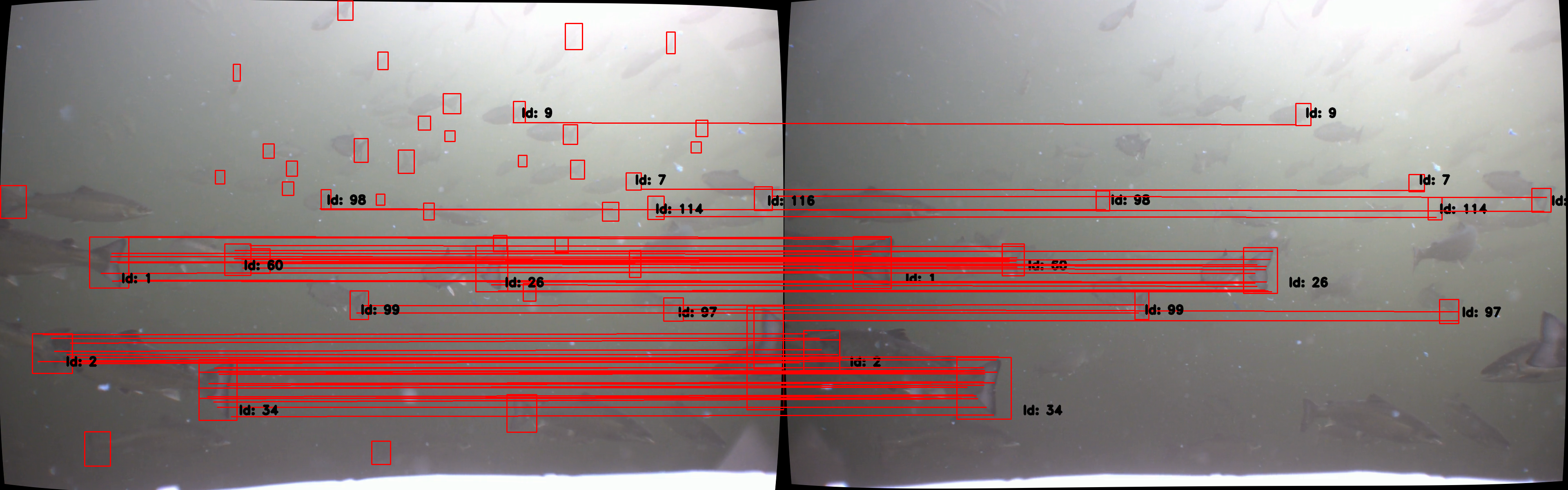}
 % \caption{Initial}
 % \label{fig:initial_img}
%\end{subfigure} % <-- added
%\begin{subfigure}[t]{0.50\textwidth}
%\centering
%  \includegraphics[scale=0.05]{Method_figures/superglue/superglue_mo_wt.png}
%  \caption{MO-WT}
%  \label{fig:mo_wt}
%\end{subfigure}\hfil % <-- added
%\begin{subfigure}[t]{\textwidth}
%\centering
%  \includegraphics[scale=0.126]{Method_figures/superglue/superglue_clahe_mo_wt.png}
 % \caption{MO-WT and CLAHE}
 % \label{fig:mo_wt_clahe}
%\end{subfigure}\hfil % <-- added
\caption{MOT and stereo matching on a single left frame with multiple different pre-processing approaches}
\label{fig:stereomatching_ex}
\end{figure*}

Figure \ref{fig:stereomatching_ex} visualizes the stereo-matching approach on a single frame.
The point correspondences are depicted when they lie within a detected bounding box on the left frame. 
\subsection{3D Multiple Object Detection and Tracking}

\begin{figure}
\includegraphics[trim={1.5cm 2.1cm 1.5cm 0cm},clip,scale=0.32]{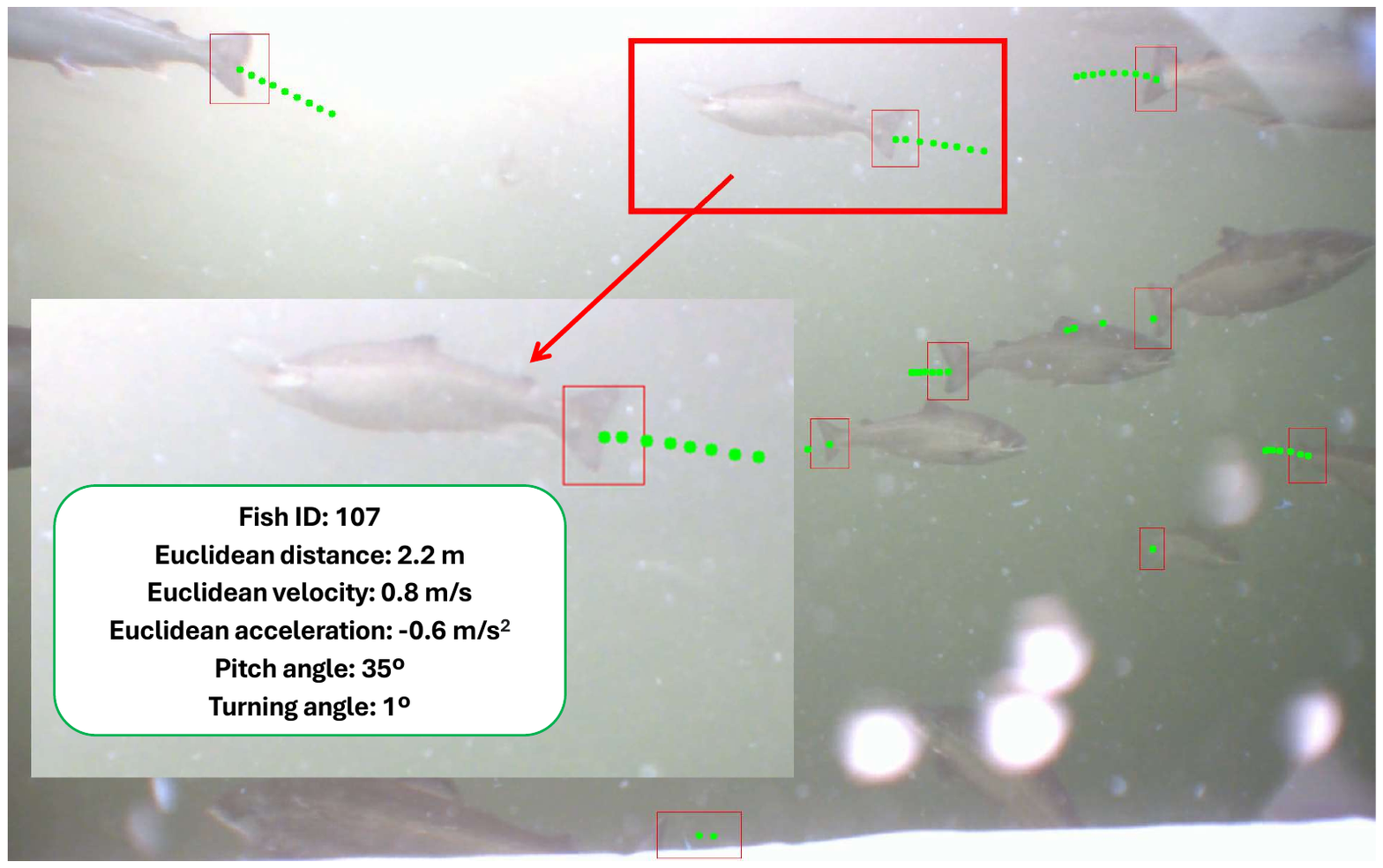}
\caption{Visualization of the tracked fish and the associated parameters of importance}
\label{fig: visualization_calculated_parameters}
\end{figure}

In the final step in the pipeline, the powers of pre-processing, YOLOv8, ByteTrack, SuperGlue, and depth estimation are finally combined to estimate the position and depth of each relevant fish in successive frames. Derived parameters can also be found from the depth- and position estimates. En example visualizing he process of tracking each fish in a frame, along with 3D position and derived parameters, is depicted in Figure \ref{fig: visualization_calculated_parameters}. Ultimately, the output is provided as the following parameters: frame number, class, fish ID, estimated fish position (x,y,z), estimated Euclidean distance, bounding box coordinates and center coordinates. 

A pseudo-algorithm depicting steps B-G in the pipeline presented in Figure \ref{fig:complete_workflow} is provided in Algorithm \ref{alg:algorithm}.

\begin{algorithm}
    \SetKwInOut{Input}{Input}
    \SetKwInOut{Output}{Output}

    \Input{Concatenated stereo image $I$, left image $I_{l}$}
    \Output{Depth estimates $x$, $y$, $z$, and $z'$}
    
    $I \leftarrow \text{Rectify}(I)$\;
    $I_{l} \leftarrow \text{Pre-process}(I_{l})$\;
    $BBs \leftarrow \text{Yolo}(I_{l})$\;
    $KP_{Matches} \leftarrow \text{Superglue}(I)$\;
    
    \If{$\text{len}(BBs) > 0$}{
        Tracks $\leftarrow \text{ByteTrack}(BBs)$\;
    }
    
    \If{Tracks and $KP_{Matches}$}{
        \ForEach{Track, BB in Tracks, BBs}{
            $KP_{masked} \leftarrow KP_{Matches}$ inside $BB$\;
            $Left_{xy} \leftarrow \text{Center}(BB)$\;
            $Diff \leftarrow Left_{xy} - \text{Mean}(KP_{masked})$\;
            $Right_{xy} \leftarrow \text{Mean}(KP_{masked r}) + Diff$\;
            
            $x, y, z \leftarrow \text{Triangulate}(Left_{xy}, Right_{xy})$\;
            $z' \leftarrow \text{RAFT-Stereo}(Left_{xy}, Right_{xy})$\;
            
            write to file $x, y, z, z'$\;
        }
    }

    \caption{Algorithm for 3D position estimation on a single frame. The following abbreviations are used: BB: bounding box, KP: key points, $x, y, z, z'$: 3D position estimates, $Left_{xy}$: Estimated center of the left image's bounding box, $Right_{xy}$: Estimated center of the right image's bounding box
    %\begin{itemize}
    %    \item BB: bounding box 
    %    \item KP: key points
    %    \item $x, y, z, z'$: 3D position estimates
    %    \item $Left_{xy}$: Estimated center of the left image's bounding box 
    %    \item $Right_{xy}$: Estimated center of the right image's bounding box 
    %\end{itemize}
    }
    \label{alg:algorithm}
\end{algorithm}

\section{Results and Discussions}
\label{sec:results}
%The proposed method outlined in Section \ref{sec:app} has been applied to the task of 3D tracking and position estimation of fish. 
%This section presents the results obtained, opening with an overview of the performance improvements achieved by using different image augmentation approaches. 
This section presents the results obtained when using the methods outlined in section~\ref{sec:app} to track fish in 3D. 
First, an an overview of the performance improvements achieved by using different image improvement approaches is provided. 
Thereafter, the results concerning the relative distance between fish and different structures from the field experiments are presented. 
Finally, new fish parameters possible to derive using the computer vision method, such as velocity and acceleration, both on an individual level and on a group basis, are presented, demonstrating the novelty and potential of this method in providing more insight into behavioral dynamics in sea-cages. 

\subsection{Image Pre-processing and Model Performance}
\subsubsection{Results}
The image pre-processing approaches outlined in Section \ref{sec:app} were tested on the training data to determine whether the pre-processing techniques had the desired effect of increasing the precision of the caudal fin detections. 
Results are presented in the form of precision, recall, mAP, and number of epochs in Table \ref{table:map_etc_image_augmentation}. 
The table is divided in two parts to separate the outputs when the methods were trained on the foreground caudal fins dataset and the full caudal fins dataset.
Labels used in the table refer to the same combinations as used earlier (i.e., "Initial", "MO-WT", "Sea-thru"), but with the addition of cases labeled "augmented" that also use the aforementioned YOLOv8n augmentation.

%The object detector is significantly good at detecting fishtails close to the camera. However, its accuracy diminishes when tasked with tracking fishtails positioned farther from the camera. This is an expected tendency due to the noise, occlusion, and low contrast, which provide adverse conditions to capture smaller tails. Additionally, human error may be more present in the dataset annotation for this scenario due to larger requirements for precision. 

All models were trained with a maximum of 300 epochs and had early stopping implemented after 50 epochs if no improvement in accuracy was detected, resulting in all models converging quickly. The mAP50 is given as the precision metric. 
The best-performing metric for each method with different image pre-processing is given in bold letters in Table \ref{table:map_etc_image_augmentation}. 

\begin{table}[h!]

\centering
\resizebox{\columnwidth}{!}
{\begin{tblr}{
    hline{1,2,3, 4,5, 6,7,8, 9, 10, 11, 12, 13, 14, 15} = {-}{},
    vline{1, 2, 3, 4, 5, 6} = {-}{}}

    \textbf{Image pre-processing} & \textbf{Precision} & \textbf{Recall} & \textbf{mAP50} & \textbf{Epochs}\\[1.3ex]
    \SetCell[c=5]{l}\textbf{Foreground caudal fins dataset}\\[1.3ex]
     Initial    & 0.886    & 0.876   & 0.945 & 154 \\[1.3ex]
    MO &  0.889  &   0.868   & 0.949 & \textbf{138} \\[1.3ex]
    MO-WT      &  0.891  &   0.888   & 0.950 & 165 \\[1.8ex]
    MO-WT + CLAHE      & \textbf{0.900}     & 0.855  & 0.949     & 156 \\[1.8ex]
    Sea-thru  & 0.882     & 0.845   & 0.933     & 154 \\[1.3ex]
    MO-WT augmented      & 0.898     & \textbf{0.895} & \textbf{0.961}    & 230 \\[1.3ex]
    \SetCell[c=5]{l}\textbf{Full caudal fins dataset}\\[1.3ex]
    Initial    & 0.815    & 0.682   & 0.775 & 134 \\[1.3ex]
    Initial augmented & \textbf{0.867} & 0.728 & 0.840 & \textbf{115} \\[1.3ex]
    MO-WT augmented &  0.841 & \textbf{0.781} & \textbf{0.860} & 132 \\[1.3ex]
    {MO-WT augmented \\ + CLAHE} & 0.857 & 0.761 &  0.849 & 131\\[1.3ex]

    \end{tblr}}
    \caption{Comparison of different image pre-processing methods, in terms of precision, recall, mAP50 and epochs}

\label{table:map_etc_image_augmentation}
\end{table}

\subsubsection{Discussion}
MO-WT augmented is, for both datasets, found to provide the best approach for object detection, both in terms of mAP50 and recall. 
Adding CLAHE has a positive effect in terms of precision, and may be beneficial if precision is more important. 
Sea-thru gave worse results than the initial method, possibly because of too little variance in colors or too much noise. 

The two datasets were found to yield differences in precision and the number of tracked and detected fish, where the Full caudal fins dataset captured a larger number of caudal fins. The precision of the object detector was however lower in this case, which can also cause more track fragmentation. More training data could mitigate this issue. 

\begin{figure*}[htbp]
 \centering
    \begin{subfigure}[t]{0.48\textwidth}
        \centering
        \includegraphics[scale=0.22, trim={0cm 1cm 0cm 1cm}, clip]{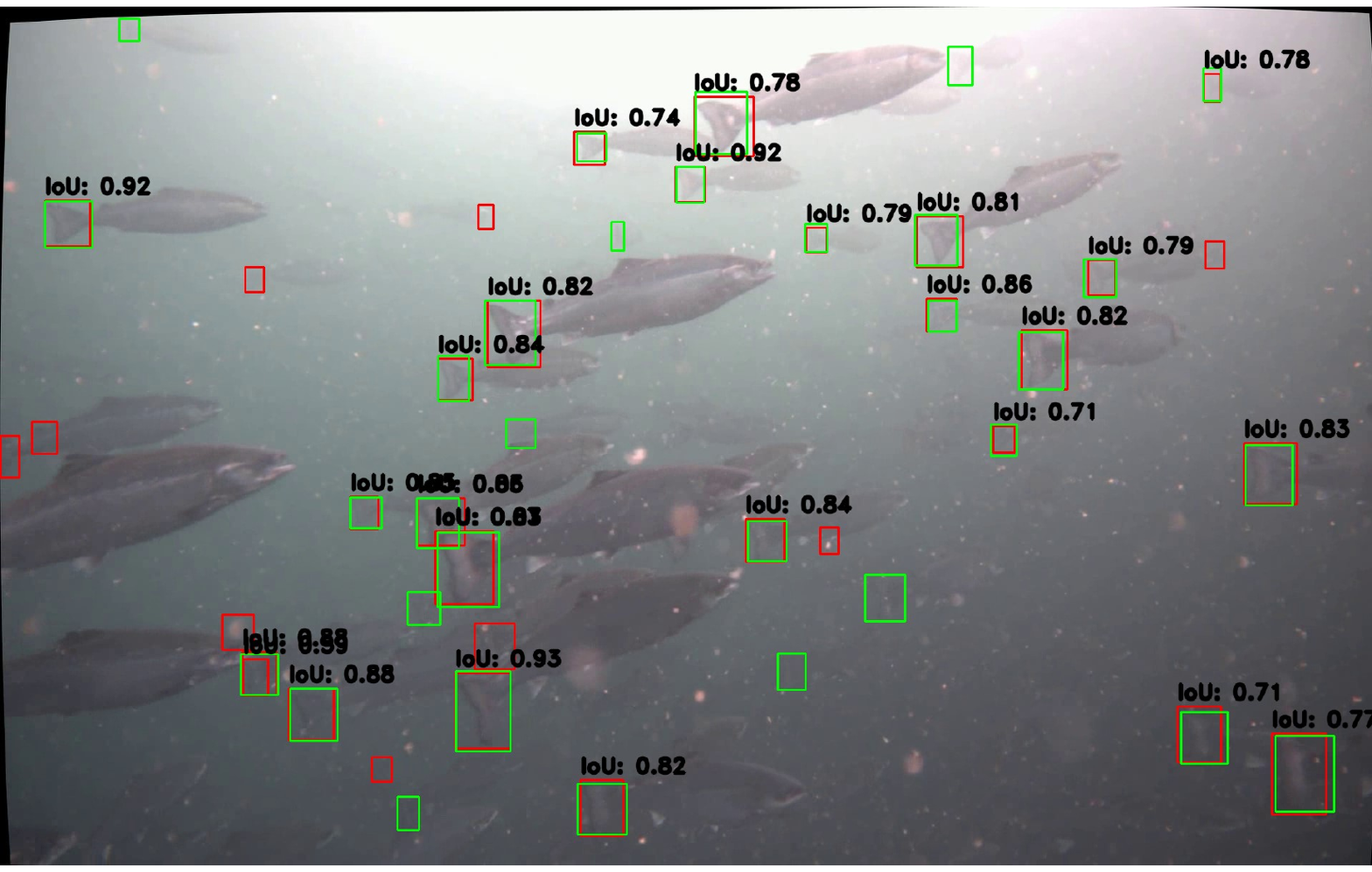}
        \caption{Initial augmented}
        \label{fig:11_nl}
    \end{subfigure}
    \begin{subfigure}[t]{0.48\textwidth}
        \centering
        \includegraphics[scale=0.22, trim={0cm 1cm 0cm 1cm}, clip]{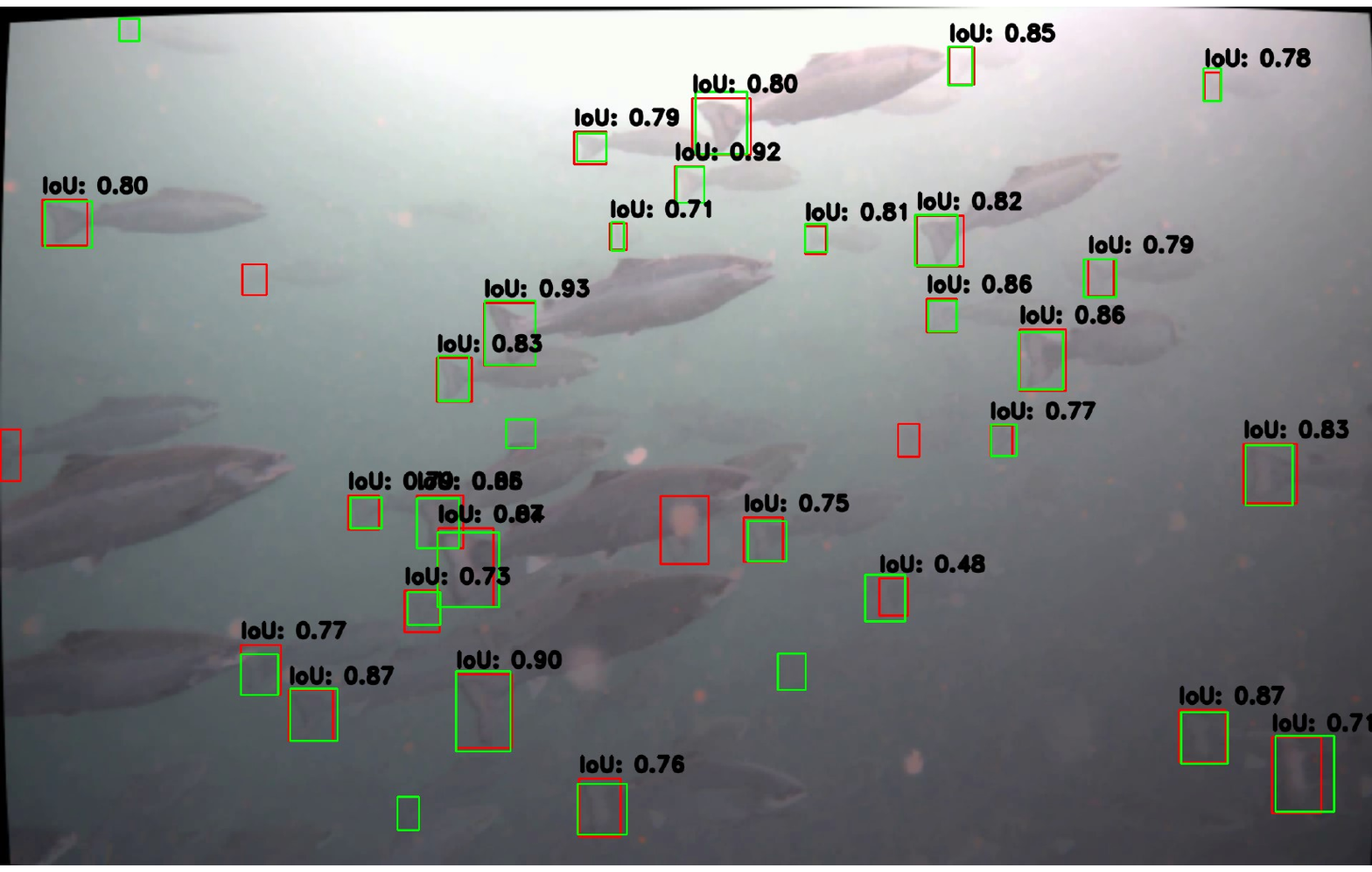}
        \caption{MO-WT}
        \label{fig:12_nl}
    \end{subfigure}
   
    \caption{A single frame of annotated and detected fishtails for two different object detection approaches. The green-colored rectangles represent the annotated caudal fins, while the red-colored rectangles represent the predicted caudal fins.}
    \label{fig:tp_fp_fn}
\end{figure*}

MO-WT augmented was, along with the initial augmented approach considered further for 3D position and tracking purposes as these two performed best.
Figure \ref{fig:tp_fp_fn} provides a visualization of these two approaches, giving an example of their object-tracking abilities. 
The visualization also illustrates the difficulties in detecting fish far away. Tails may become increasingly difficult to track, and the accuracy of the bounding box estimate may become worse.
While the object detector is good at detecting caudal fins on fish close to the camera, its accuracy diminishes when tasked with tracking fish positioned further from the camera. 
This is not unexpected since challenges such as image noise, occlusion, and low contrast are more prominent over long distances, and the fins appear smaller in the image for fish that are far away. These factors can also increase the chances of human error during dataset annotation as this is also rendered more difficult by adverse conditions. 

\subsection{Trajectory Smoothing}
\subsubsection{Results}
\begin{figure}
    \centering % <-- added
    
\begin{subfigure}[t]{0.48\textwidth}
  \includegraphics[width=\linewidth, trim={3cm 5cm 4cm 4.5cm},clip]{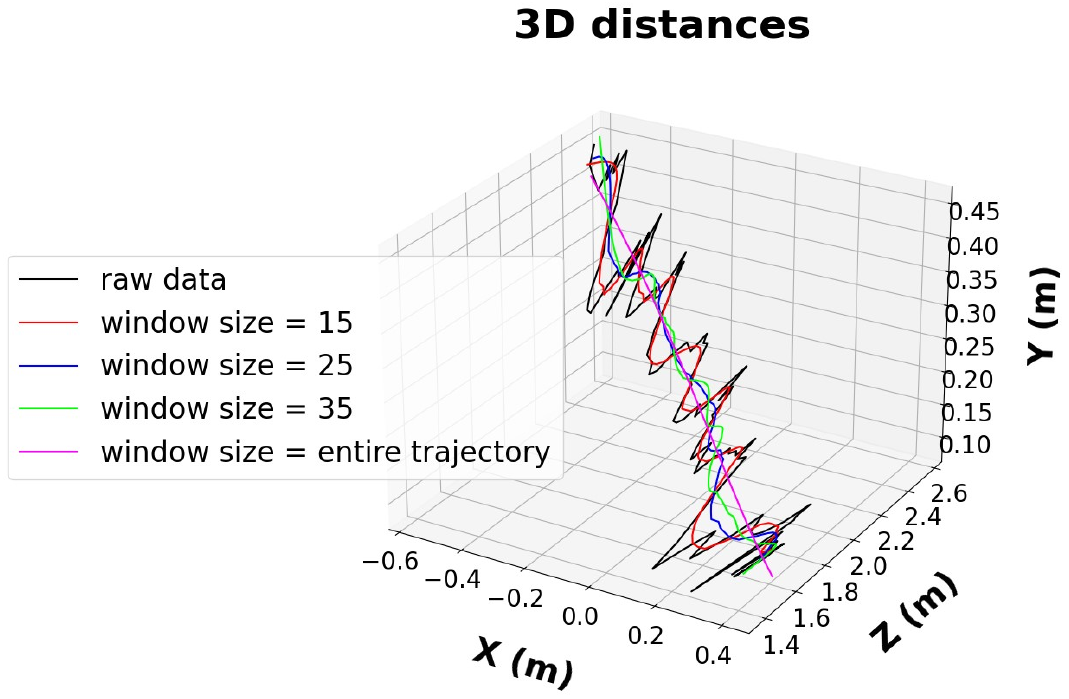}
  \label{fig:sg_3d_position}
\end{subfigure}\hfil % <-- added
\begin{subfigure}[t]{0.48\textwidth}
  \includegraphics[trim={0cm 4cm 0cm 4.5cm},clip,width=\linewidth]{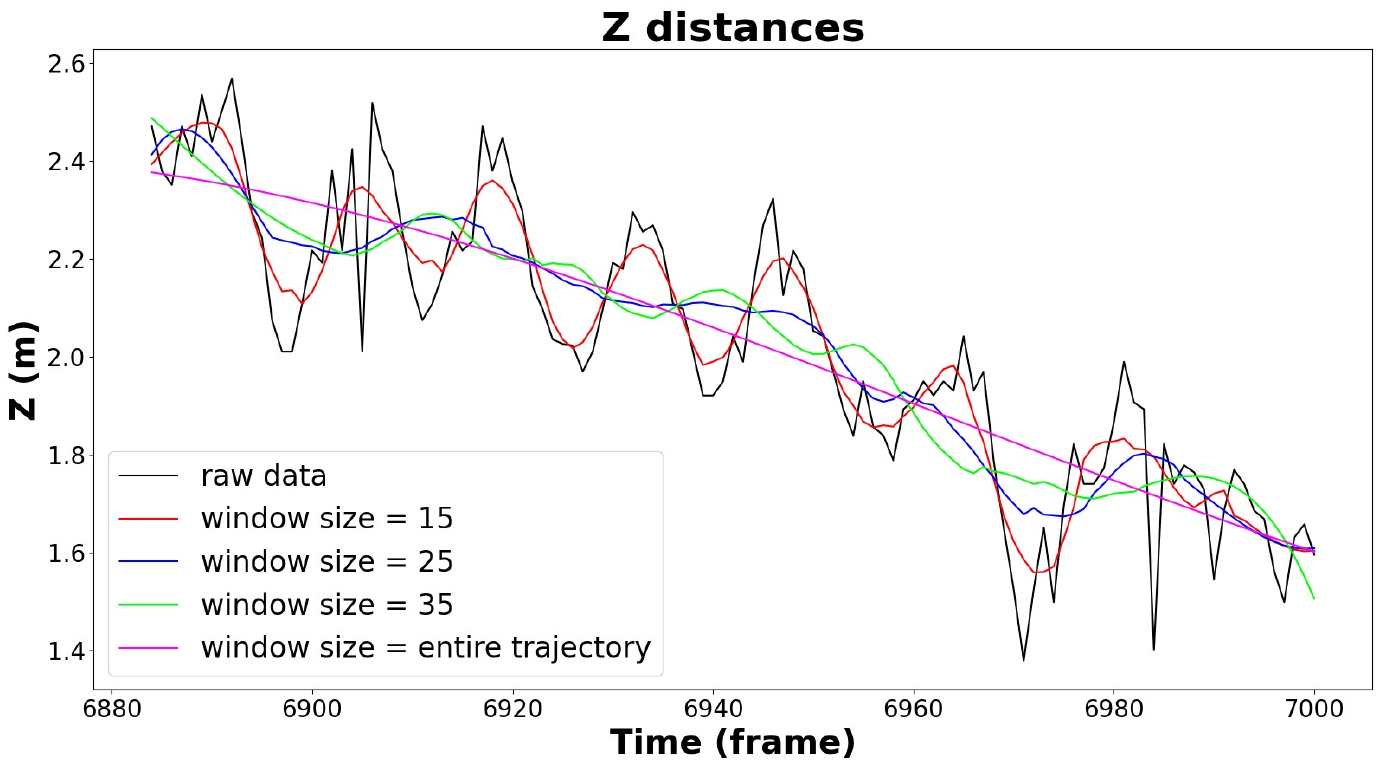}
  \label{fig:sg_z_positions}
\end{subfigure}\hfil % <-- added
\begin{subfigure}[t]{0.48\textwidth}
  \includegraphics[trim={0cm 4cm 0cm 4cm},clip,width=\linewidth]{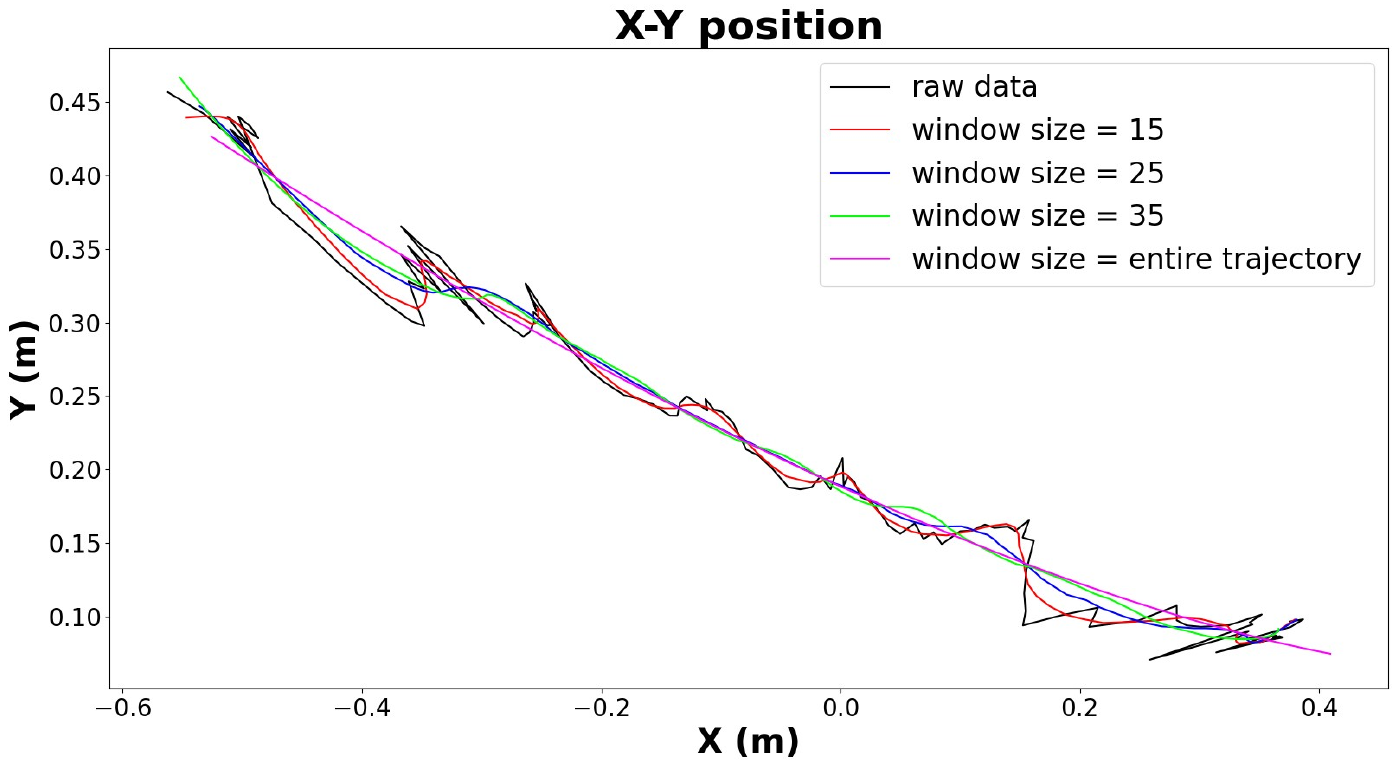}
  \label{fig:sg_xy_position}
\end{subfigure}\hfil % <-- added
\caption{Comparison of different window lengths concerning a Savitzky-Golay smoothing approach for the trajectory of a single fish}
\label{fig: SG_filtering}
\end{figure}

The estimated 3D position values of individual fish were found to fluctuate considerably across subsequent frames. 
This may be due to differences in the accuracy of the center coordinates of the fish tails in the left and right frame. 
Furthermore, the dynamic movement of the tracked caudal fins due to tail beat action \cite{warren2023novel} presents an additional challenge, potentially causing less reliable estimations. 
The Savitzky-Golay (SG) filter \cite{savgol} smoothed out the trajectories by fitting the trajectory to a third-order polynomial using the least squares approach. 
Figure \ref{fig: SG_filtering} demonstrates how the difference in window sizes (i.e., the number of frames used in each segment approximated by a polynomial) causes variations while still keeping the same overall pattern. 
Window sizes spanning the entire trajectory or 15, 25, or 35 frames were tested.
\subsubsection{Discussion}
While the trajectories for raw and data small window sizes fluctuated more than for larger window sizes, all approaches had similar mean values, implying that they captured the distance to the fish. 
The periodic fluctuations were expected as tracking of the caudal fin would pick up the tail beats of the fish while swimming. 
Tracking the overall path of the fish, as was the intention here, woudl thus be best achieved using the largest window sizes. 
%In this scenario, the tail beat-induced movement was evident from the sinusoidal motion observed in the graph.
%To reduce the impact of this disturbance, a window size spanning the entire trajectory was used, yielding improved tracking of the overall path since this was the main focus in the present study.
In other scenarios where tail beats are of interest, it could conversely be of interest to accentuate these variations to thereby derive the tail beat frequency and possibly amplitude, which are believed to be good indicators of swimming activity in salmon \cite{warren2023novel}.

\subsection{Depth Estimation}
\subsubsection{Results}
\begin{figure}[h!]
    \centering % <-- added
    
\begin{subfigure}[t]{0.48\textwidth}
  \includegraphics[width=\linewidth]{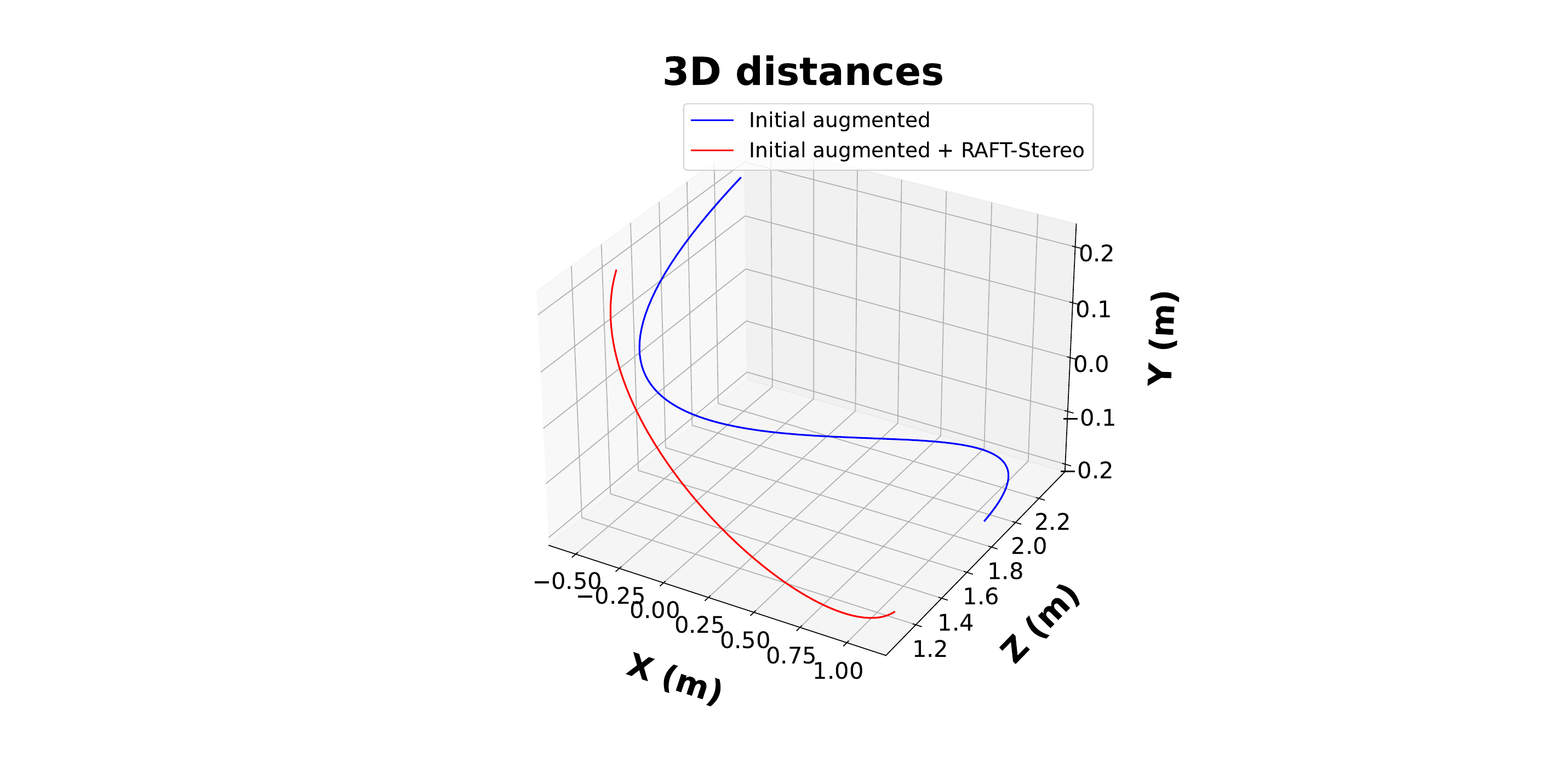}
  \label{fig:raft_1}
\end{subfigure}\hfil % <-- added
\begin{subfigure}[t]{0.48\textwidth}
  \includegraphics[width=\linewidth]{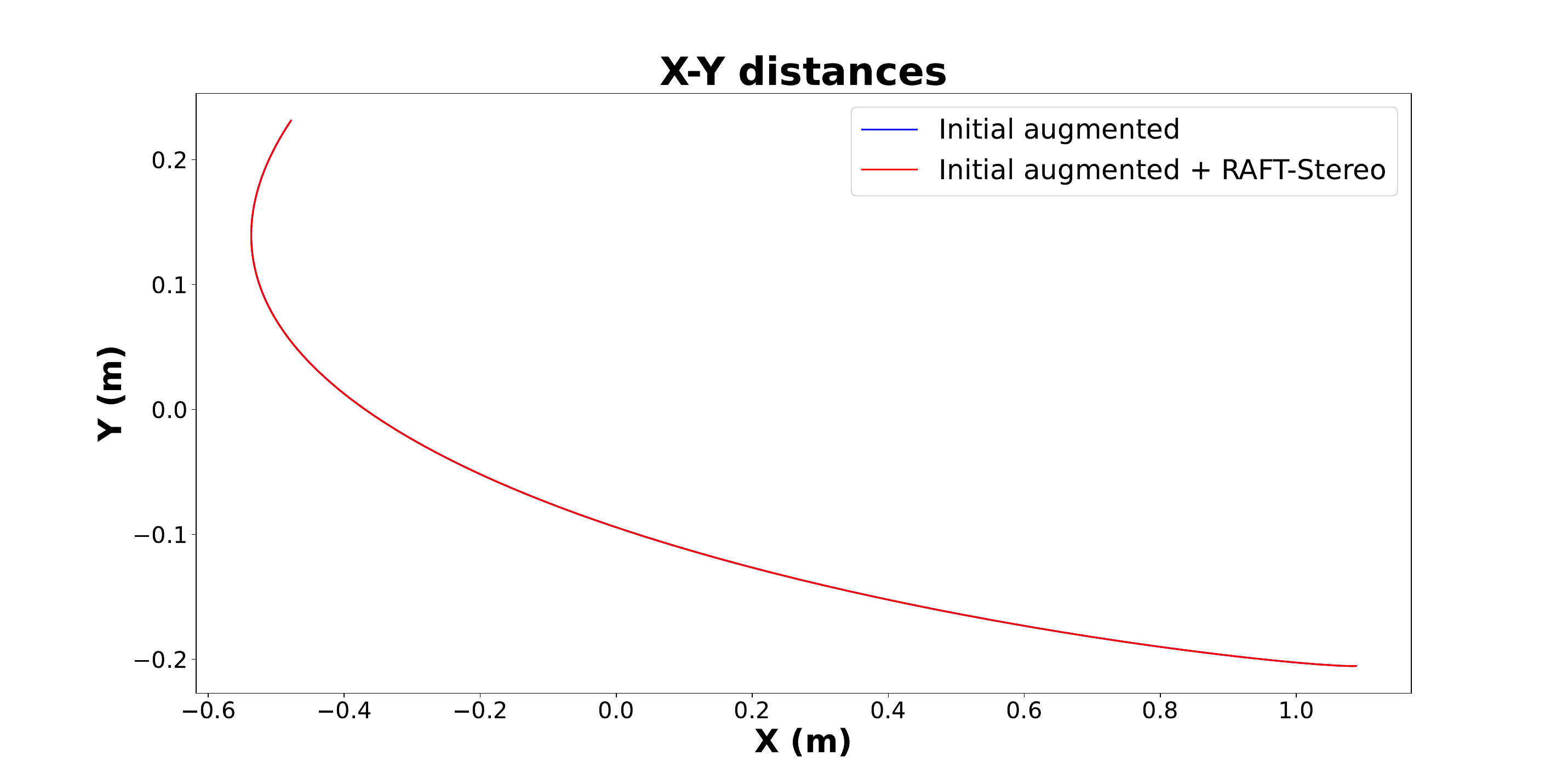}
  \label{fig:raft_1}
\end{subfigure}\hfil % <-- added
\begin{subfigure}[t]{0.48\textwidth}
  \includegraphics[width=\linewidth]{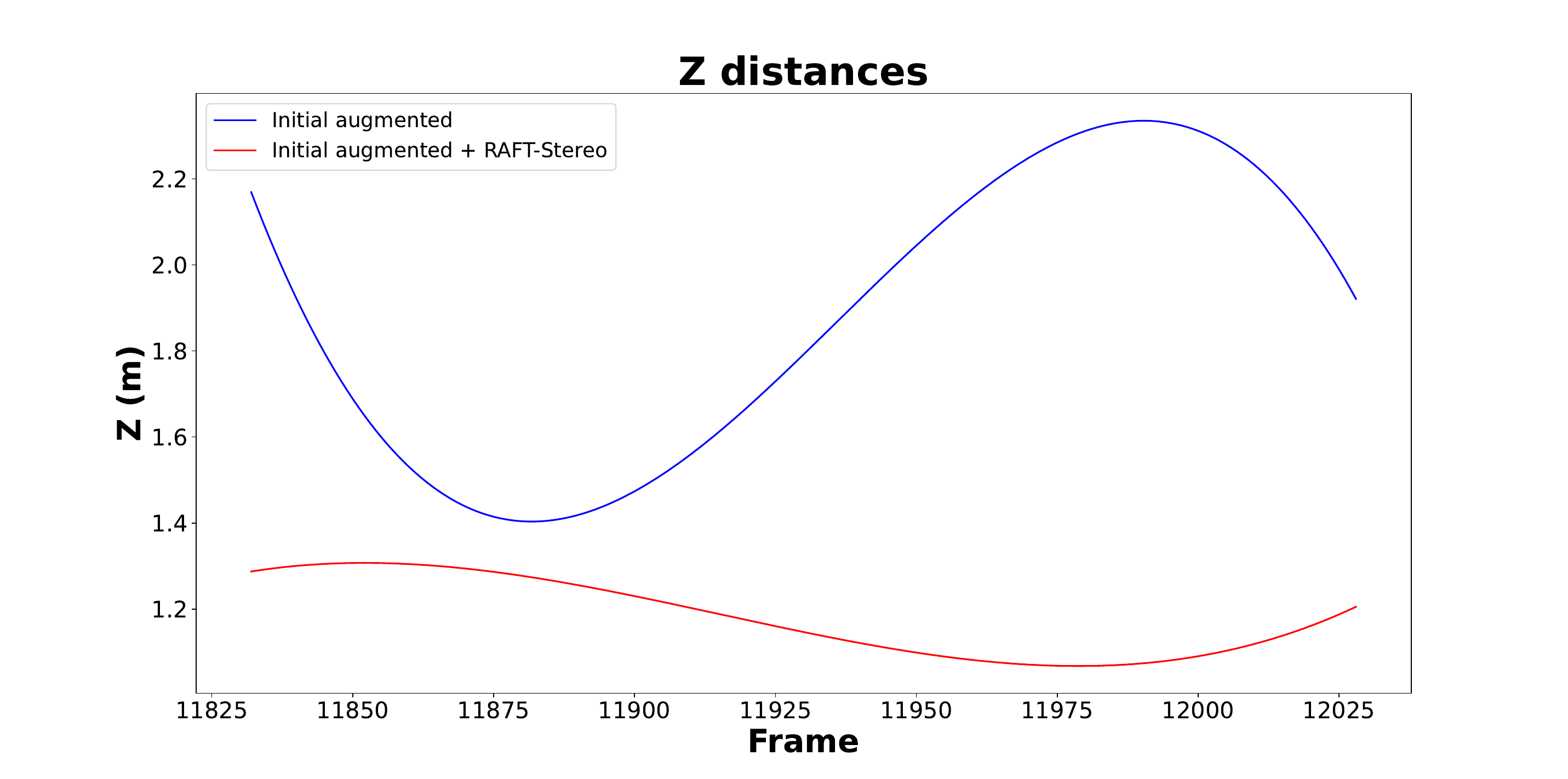}
  \label{fig:raft_2}
\end{subfigure}\hfil % <-- added
\caption{Comparison of RAFT-Stereo and classical triangulation on the trajectory of a single fish}
\label{fig: RAFT-Stereo_comparison}
\end{figure}

Two depth estimation approaches were tested; RAFT-Stereo and classical triangulation. 
Figure \ref{fig: RAFT-Stereo_comparison} shows data from both methods describing an individual salmon visually observed to exhibit a behavior where it was first moving away from the structure and then approaching it at the end of the trajectory. 
The cases shown are for the smoothed initial augmented approach with triangulation and the smoothed initial augmented approach with RAFT-stereo included. 

\subsubsection{Discussion}
Classical triangulation proved more reliable than RAFT-stereo, as it provided outputs that matched the visual observations. 
RAFT-Stereo did not capture this behavior, instead apparently depicting a type of behavior where the fish approaches the camera slightly, thereafter keeping a somewhat constant track away from it. 
This implies that RAFT-Stereo may be less robust than triangulation, and may be attributed to a lack of training in underwater training data. 
\subsection{Analysis of Distance to Varying Intrusive Objects}

\subsubsection{Results}
\begin{figure*}
    \centering
    \begin{subfigure}[t]{1\textwidth}
        \centering
        \begin{minipage}{0.3\textwidth}
            \includegraphics[width=\linewidth]{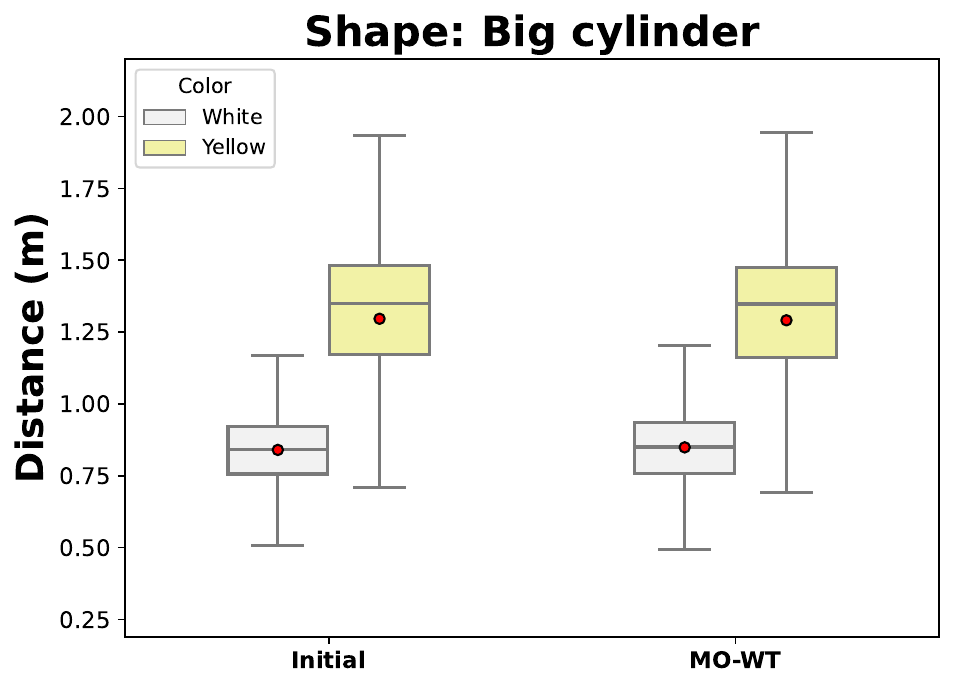}
            \caption{Big cylinder}
            \label{fig:big_cylinder}
        \end{minipage}
        \begin{minipage}{0.3\textwidth}
            \includegraphics[width=\linewidth]{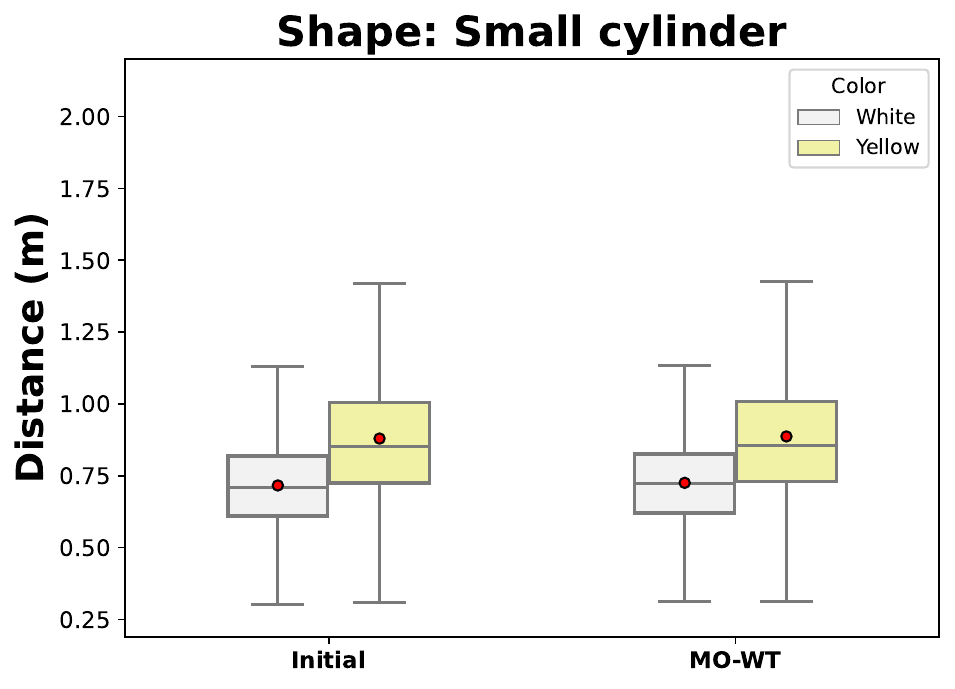}
            \caption{Small cylinder}
            \label{fig:small_cylinder}
        \end{minipage}

        \begin{minipage}
            {0.3\textwidth}
            \includegraphics[width=\linewidth]{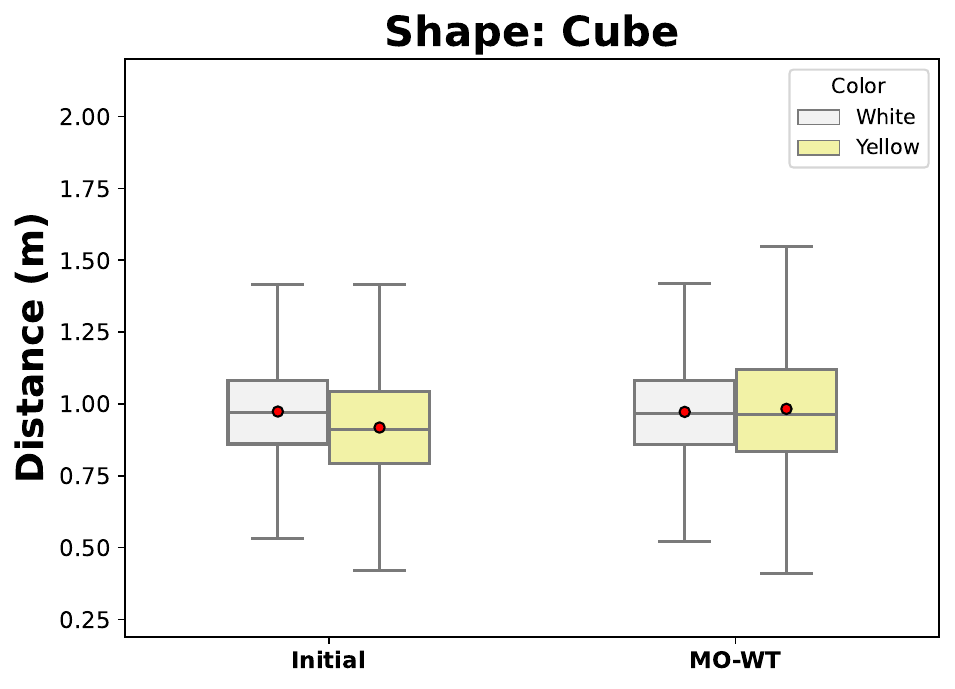}
            \caption{Cube}
            \label{fig:cube}
        \end{minipage}
    \end{subfigure}
    
    \caption{Boxplots comparing the average aggregated minimum distance estimates between tracked fish and structures for the two object detection methods trained on the full caudal fins dataset, Initial augmented and MO-WT augmented}
    \label{fig:boxplots_shapes}
\end{figure*}

For each frame in the videos, the 3D distance position estimates were gathered, and out of these, the closest value based on the Euclidean distance was extracted. The closest values for each frame for a given shape and structure were then extracted and aggregated. 
The mean and standard deviations of the minimum distance kept to each structure size, shape, and color are shown in Table \ref{table:aggregated_values_minimum}, while Table \ref{table:mean_agg_distance} shows the average distances between fish and structures for the same cases. 
Boxplots visualize the distribution in Figure \ref{fig:boxplots_shapes} for each shape and in Figure \ref{fig:boxplots_colors} for each color. Further results for distance estimations and differences between the methods in terms of quantity of detected and tracked fish are found in Appendix 1 and Appendix 2. 

Based on the presented results and those in Appendix 1 and Appendix 2, an overview of the features of each model with capabilies is shown in Table \ref{table:comparison_models_checkmarks}.

\subsubsection{Discussion}
Regarding the size of the structures, the table and figures provide a quite clear indicator that fish stay closer to smaller objects than larger objects. From Table \ref{table:aggregated_values_minimum}, the aggregated minimum average distance between the closest fish and each structure shows clearly that the the fish keep closer to the small cylinder, than to the larger structures, the cube and the big cylinder. This also seems to be the case in general, based on the estimates of the average distance the tracked fish keep to the structures in Table \ref{table:mean_agg_distance}. 

%\subsubsection{Color}

\begin{figure*}
    \centering
    \begin{subfigure}[t]{1\textwidth}
        \centering
        \begin{minipage}{0.30\textwidth}
            \includegraphics[width=\linewidth]{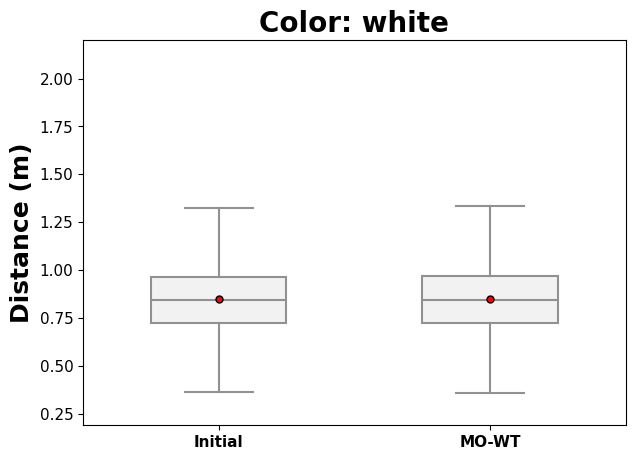}
            \caption{White}
            \label{boxplots_withe}
        \end{minipage}
        \begin{minipage}{0.30\textwidth}
            \includegraphics[width=\linewidth]{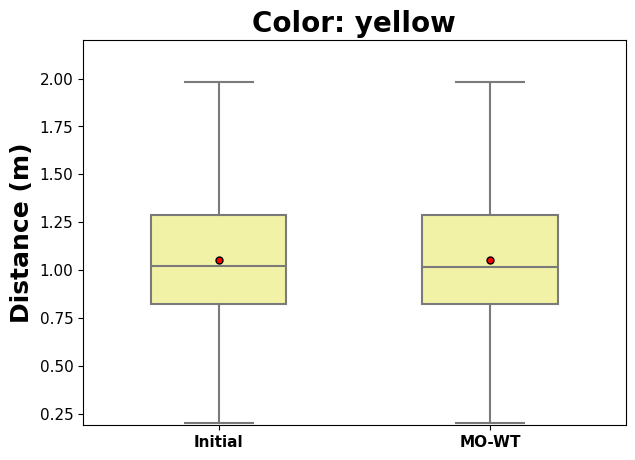}
            \caption{Yellow}
            \label{fig:boxplots_yellow}
        \end{minipage}
    \end{subfigure}
    
    \caption{Boxplots comparing the aggregated average minimum distance estimates between tracked fish and colors for the two object detection methods trained on the full caudal fins dataset, Initial augmented and MO-WT augmented}
    \label{fig:boxplots_colors}
\end{figure*}

The boxplots in Figure \ref{fig:boxplots_colors} show that the fish stay closer to the white structures than the yellow structures. This also seems to be the case in Table \ref{table:aggregated_values_minimum}, although the distance the fish keep to the white cube compared to the yellow cube is not as prominent, and the conclusion regarding which color the fish is more drawn to is different for the initial augmented method than for the other methods. Table \ref{table:mean_agg_distance} nonetheless shows a clear trend of the fish, in general, staying closer to the white than to the yellow cube, so the consensus seems to be that fish stay closer to white structures. 
%The degree of dispersion is higher for the yellow structures than for the white structures, indicating greater differences in individual fish responses when faced with yellow structures. The enlargement of the standard deviation may also happen due to a decrease in depth estimation accuracy when the closest fish is positioned further away from the stereo camera system. 

%\subsubsection{Shape}

Results were less conclusive on whether or not the fish had a preference in shape. The average distances kept by the fish to the cube and the big cylinder were varying, as shown in Table \ref{table:aggregated_values_minimum}.

\begin{table}[ht!]
\resizebox{\columnwidth}{!}
{\begin{tblr}{
  vline{1, 3, 5, 8, 9} = {-}{},
  hline{2,3} = {3-8}{},
  hline{1,4,6,8,10, 13, 16} = {-}{},
  cell{1}{3,4} = {c=3}{c},
}
 Case & Color  &  \textbf{Average minimum distance [m]} &\\
 & & Full caudal fins dataset & & Foreground caudal fins dataset & & \\
  & & {Initial augmented} & {MO-WT \\ augmented} & Initial & {MO-WT} & {MO-WT + \\ RAFT-Stereo}\\
 Big Cylinder & White  & {0.84} $\pm$ {0.13} & 0.85 $\pm$ 0.14 & {0.85} $\pm$ 0.13 & {0.85} $\pm$ 0.14 & 0.87 $\pm$ {0.10} \\
 & Yellow &  {1.30} $\pm$ {0.29} & 1.29 $\pm$ {0.29} & {1.15} $\pm$ 0.31 & 1.04 $\pm$ 0.27 & 0.98 $\pm$ {0.11} \\
 Small Cylinder & White &  {0.72} $\pm$ {0.15} & 0.73 $\pm$ 0.16 & 0.80 $\pm$ 0.15 & 0.82 $\pm$ 0.16  & {0.79} $\pm$ {0.13} \\
 & Yellow  & 0.88  $\pm$ {0.24} & {0.89} $\pm$ {0.24} &  0.93 $\pm$ 0.25 & {0.92} $\pm$ 0.25 & 1.09 $\pm$ {0.22} \\
 Cube & White & 0.97 $\pm$ {0.17} &  0.97 $\pm$ 0.18 &   0.94 $\pm$ 0.18 & {0.92}  $\pm$ {0.16} & 0.82 $\pm$ 0.21\\
 & Yellow & 0.92 $\pm$ {0.19}  & {0.98} $\pm$ 0.23 & 0.95 $\pm$  0.20 & {0.99} $\pm$ 0.22  & 0.75 $\pm$ {0.16} \\

\end{tblr}}
\caption{Comparison of different methods, in terms of minimum average aggregated distances between the fish and the camera}
\label{table:aggregated_values_minimum}
\end{table}

\begin{table}
\centering

\resizebox{\columnwidth}{!}
{\begin{tblr}{
  vline{1, 3, 7} = {-}{},
  vline{5} = {2-10}{},
  hline{2} = {3-6}{},
  hline{1,3,4, 6,8,10} = {-}{},
  cell{1}{3,4} = {c=3}{c},
}
 \textbf{Shape} & \textbf{Color} & \textbf{Average distance [m]}\\
  & & Full caudal fins dataset & & Foreground caudal fins dataset \\
  &  & {Initial augmented}  & {MO-WT augmented} & Initial & {MO-WT}  \\
  
 Big Cylinder & White & 1.85 & 1.85 & 1.84 & 1.79 \\
 & Yellow  & 2.13 & 2.13 & 2.07 & 1.96 \\
  Cube & White & 1.73 & 1.73 & 1.62 & 1.60 \\
 & Yellow  & 1.91 &1.91 & 1.83 & 1.84 \\
 Small Cylinder & White & 1.68 & 1.68 & 1.62 & 1.63 \\
 & Yellow & 1.79 & 1.79 &  1.66 & 1.67 

\end{tblr}}
\caption{Comparison of methods in terms of average distances between the fish and the camera}

\label{table:mean_agg_distance}
\end{table}

% \begin{table}[hbt!]
% \centering

% \resizebox{0.487\textwidth}{!}
% {
% \begin{tblr}{
%   vline{1,2,3,4,5,6, 7, 8 ,9} = {-}{},
%   hline{1,2, 3,4, 5,6, 7,8,9} = {-}{},
% }
%   \textbf{Dataset} & \textbf{Model} & \textbf{Fast} & \textbf{{Tracks \\ fish}} & \textbf{{Detects fish \\
%   further away}} & \textbf{{Most \\ tracks}} \\
%    Foreground caudal fins dataset & Initial & \cmark & \cmark & &   \\
%    & MO-WT augmented & & \cmark  &  &  \\
%     Full caudal fins dataset & MO-WT augmented &  & \cmark & \cmark &   \\
%     & Initial augmented & \cmark & \cmark & \cmark & \cmark\\
% \end{tblr}

% }
% \caption{Performance evaluation of the proposed methods}
%  \label{table:comparison_models_checkmarks}
% \end{table}

\begin{table}[hbt!]
\centering
\resizebox{0.487\textwidth}{!}{
\begin{tabular}{l l c c c c}
\hline
\textbf{Dataset} & \textbf{Model} & \textbf{Fast} &
\makecell{\textbf{Tracks} \\ \textbf{fish}} &
\makecell{\textbf{Detects fish} \\ \textbf{further away}} &
\makecell{\textbf{Most} \\ \textbf{tracks}} \\
\hline

Foreground caudal fins dataset & Initial & \cmark & \cmark & & \\
Foreground caudal fins dataset & MO-WT augmented & & \cmark & & \\
Full caudal fins dataset & MO-WT augmented & & \cmark & \cmark & \\
Full caudal fins dataset & Initial augmented & \cmark & \cmark & \cmark & \cmark \\
\hline
\end{tabular}
}
\caption{Performance evaluation of the proposed methods}
\label{table:comparison_models_checkmarks}
\end{table}

% \begin{table}[hbt!]
% \centering

% \resizebox{0.487\textwidth}{!}
% {\begin{tblr}{
%   colspec = {llllll},
%   vlines,
%   hlines,
% }
% \textbf{Dataset} &
% \textbf{Model} &
% \textbf{Fast} &
% \textbf{Tracks \\ fish} &
% \textbf{Detects fish \\ further away} &
% \textbf{Most \\ tracks} \\

% Foreground caudal fins dataset & Initial & yes & yes &  &  \\
% Foreground caudal fins dataset & MO-WT augmented &  & yes &  &  \\

% Full caudal fins dataset & MO-WT augmented &  & yes & yes &  \\
% Full caudal fins dataset & Initial augmented & yes & yes & yes & yes
% \end{tblr}}
% \caption{Performance evaluation of the proposed methods}
% \label{table:comparison_models_checkmarks}
% \end{table}

%\subsubsection{Discussion}
The results concerning the sizes, shapes and colors align with those concluded in \cite{sonar_data}. 
The fish preferring to keep larger distances to larger objects is in accordance with expected behaviors from salmon, as a larger object may occur as a larger collision risk than smaller objects. 
Similarly, the propensity to stay further from yellow than white objects confirmed previous observations made using sonars \cite{sonar_data}. 
The greater dispersion in distances kept to yellow than to white structures was an interesting observation that could indicate greater differences in individual fish responses when faced with yellow structures. 
This increase in standard deviation could also happen due to a decrease in depth estimation accuracy when the closest fish is positioned further away from the stereo camera system. 
Unlike for size and color, the results were inconclusive with respect to the impact of structure shape.  

%The results concerning the sizes, shapes and colors align with those concluded in \cite{sonar_data}. 
As for the 3D distance estimation, both MO-WT augmented and Initial augmented provide image qualities enabling the estimation of this distance.% and reach the same conclusion. 
MO-WT augmented has a higher mAP50 and recall on the annotated dataset, while Initial and Initial augmented has a higher precision and a higher number of FPS (2.7, 13.1 and 12.8 respectively)  due to skipping the pre-processing step, which is also reflected in Table \ref{table:comparison_models_checkmarks}. The closed-loop control system of an average UUV utilized in an underwater industrial fish farming environment operates at approximately 10 Hz. The initial approaches operate faster than this, but this is not the case for MO-WT augmented. Additional average FPS values for each method can be found in Appendix 1, depicting a clear relation between the utilization of pre-processing and a higher inference time. This shows that there is a clear trade-off between the pre-processing approaches and the approaches not involving pre-processing. By utilizing GPU-accelerated methods, it is thought that this trade-off can be mitigated. 
%By combining the table values in this section and the appendices, the initial augmented approach and MO-WT augmented approach seem to be the best, as they track a similar amount of fish in practice. Initial augmented can however be found to track a larger amount in all cases apart from the yellow big cylinder, and is therefore deemed the best approach at present. While MO-WT should still be considered, it may therefore not necessarily be favored, which is reflected in Table \ref{table:comparison_models_checkmarks}. 
\subsection{Derived Parameters: Fish Behaviour Metrics}
The behavior of fish can be described and assessed by different metrics, including position, velocity, acceleration, turning angle, and pitch angle. 
While the usefulness of position (and hence distance) as an indicator is outlined by the results presented earlier, the full utility of the method would be better demonstrated by presenting similar data on the other four metrics. 
This section therefore presents these four parameters computed from the same field dataset as used above. 
%These five metrics will be considered in the result section. 
All metrics are relative to the camera frames and are expressed based on the unit of meters (m), while some also contain seconds (s). 
The left camera marks the origin of the real 3D coordinate frame. While the different positions ($p_x$, $p_y$, $p_z$) are provided by triangulation, the Euclidean distance is given by:
\begin{equation}
    euclidean_{dist} = \sqrt{p_x^2+p_y^2+p_z^2}
    \label{eq:eucledian_dist}
\end{equation}
When the frame rate given in frames per second (FPS) is known, it can be used to calculate the velocities ($v_x$, $v_y$ and $v_z$) and the accelerations ($a_x$, $a_y$ and $a_z$):
\begin{equation}
    v_{i,j} = (p_{i,j} - p_{i, j-1})FPS
    \label{eq:vel}
\end{equation}
\begin{equation}
    a_{i,j} = (v_{i, j} - v_{i, j-1})FPS
    \label{eq:acc}
\end{equation}
where $p$ and $v$ represent the positions and velocities relative to the camera system, respectively, $i$ serves as a placeholder for the calculation in dimensions $x$, $y$, or $z$, and $j$ represents the timestep given in the number of frames.

The turning angle and the pitch angle are slightly more complex. 
The turning angle is found by considering the velocity vector of a fish, and estimates the degree of motion changes, thus representing the change in angular direction between these two timesteps. 
The calculation of the turning angle is obtained by:
\begin{equation}
\phi_i = \arccos{\frac{\mathbf{v}_j \cdot \mathbf{v}_{j-1}}{\| \mathbf{v}_j \|\| \mathbf{v}_{j-1} \|}}
    \label{eq:turning_angle}
\end{equation}
where $\mathbf{v}_{j-1}$ and $\mathbf{v}_j$ are the 3D velocity vectors in two consecutive timesteps.  

The pitch angle, on the other hand, represents how the fish moves towards or away from the camera in each timestep. Unlike the turning angle, it considers the position of the fish and does not provide a change in angles. 
The formula used in this study to obtain the pitch angle is given by: 
\begin{equation}
\theta = \arctan \frac{\sqrt{(p_z^{i} - p_z^{i-1})^2}}{\sqrt{(p_x^{i} - p_x^{i-1})^2 + (p_y^{i} - p_y^{i-1})^2}}
    \label{eq:pitch_angle}
\end{equation}
\subsubsection{Results}
Velocity, acceleration, pitch angle, and turning angle were derived for all trakced fish in the datatsets for the different shapes, colors and sizes, and subsequently averaged across individuals for each case (Table \ref{table:derived_parameters})
%Velocity, acceleration, pitch angle, and turning angle were derived from the estimated 3D positions of the fish. The process of tracking each fish in a frame, along with 3D position and derived parameters can yield an estimate for several fish, which may be interesting when considering fish schooling behavior. 
% The derivation of such parameters may provide interesting knowledge about both the fish schooling behavior and the individual behavior of each fish.
%Table \ref{table:derived_parameters} provides the average values of the derived parameters for each intrusive object. 
%In the case of the reactions to the different intrusive objects, the average values are low for each parameter, but with 
Velocity was higher for yellow structures than white ones. 
Otherwise, computed values did not vary much across color, size or type of structure. The pitch value was consistently low, suggesting that fish keep a constant distance away from the object. 
Further results are given in Appendix 2.

\begin{figure*}[h!]
    \centering
    \begin{subfigure}[t]{1\textwidth}
        \centering
        \begin{minipage}{0.40\textwidth}
            \includegraphics[trim={0cm 4cm 0cm 4.5cm},clip,width=\linewidth]{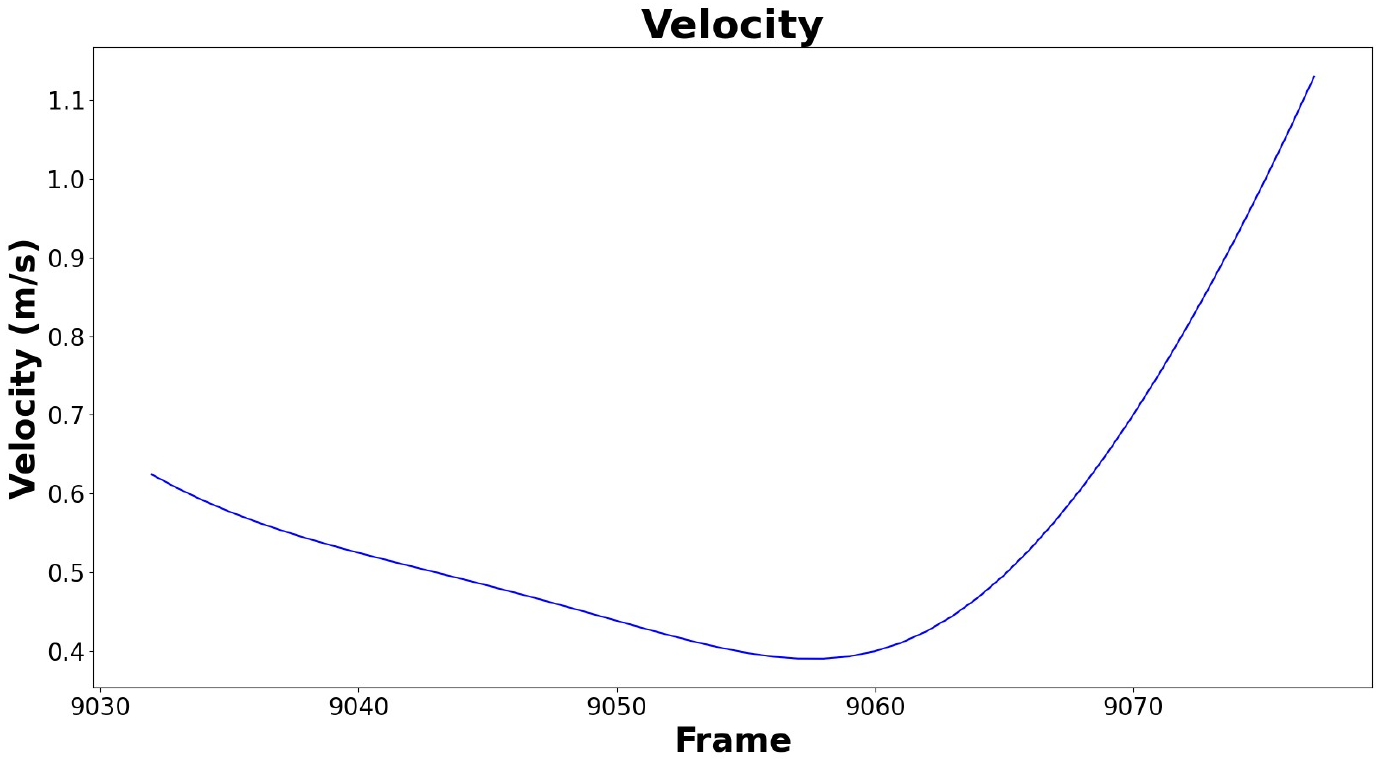}
            \caption{Velocity}
            \label{fig:velocity}
        \end{minipage}
        \begin{minipage}{0.40\textwidth}
            \includegraphics[trim={0cm 4cm 0cm 4.5cm},clip,width=\linewidth]{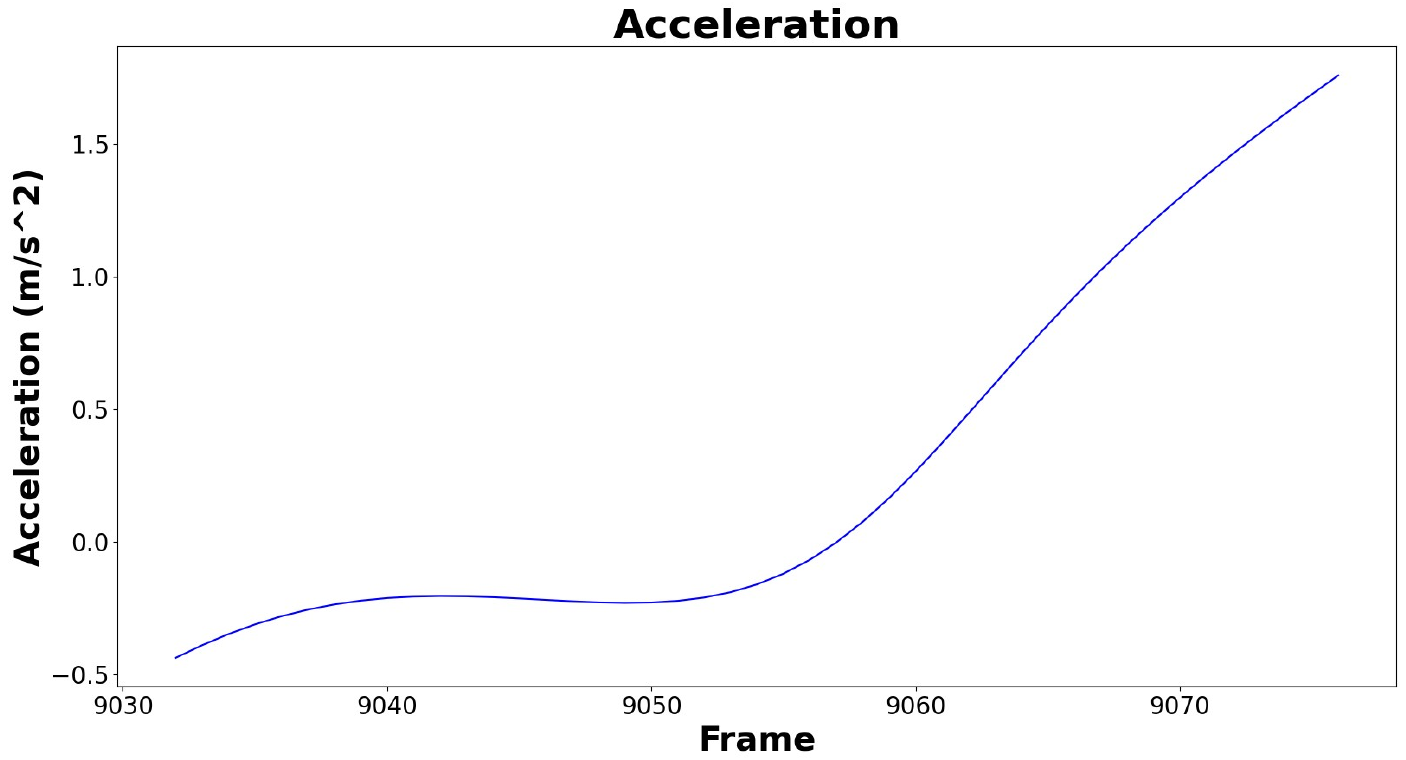}
            \caption{Acceleration}
            \label{fig:acceleration}
        \end{minipage}

        \begin{minipage}{0.40\textwidth}
            \includegraphics[trim={0cm 4cm 0cm 4.5cm},clip,width=\linewidth]{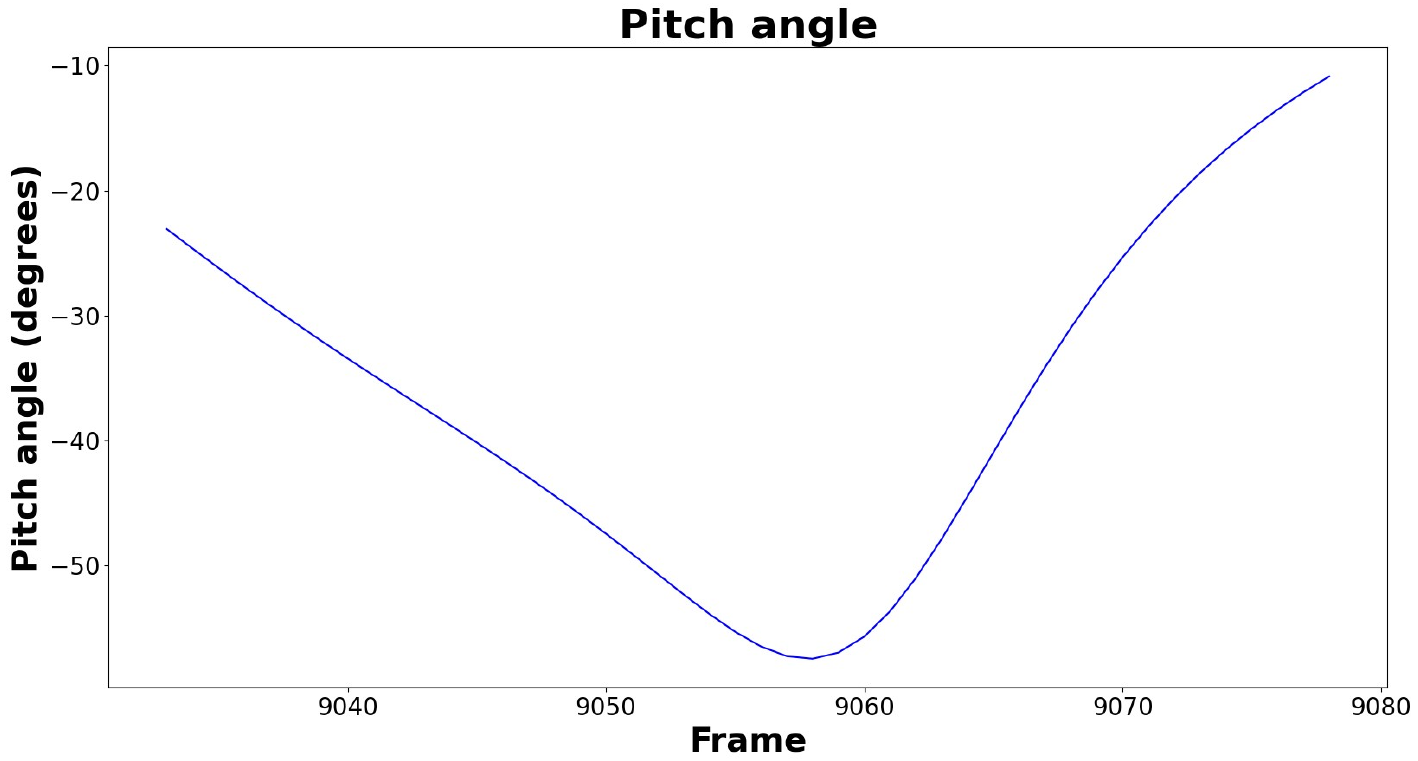}
            \caption{Pitch angle}
            \label{fig:pitch_angle}
        \end{minipage}
        \begin{minipage}{0.40\textwidth}
            \includegraphics[trim={0cm 4cm 0cm 4.5cm},clip,width=\linewidth]{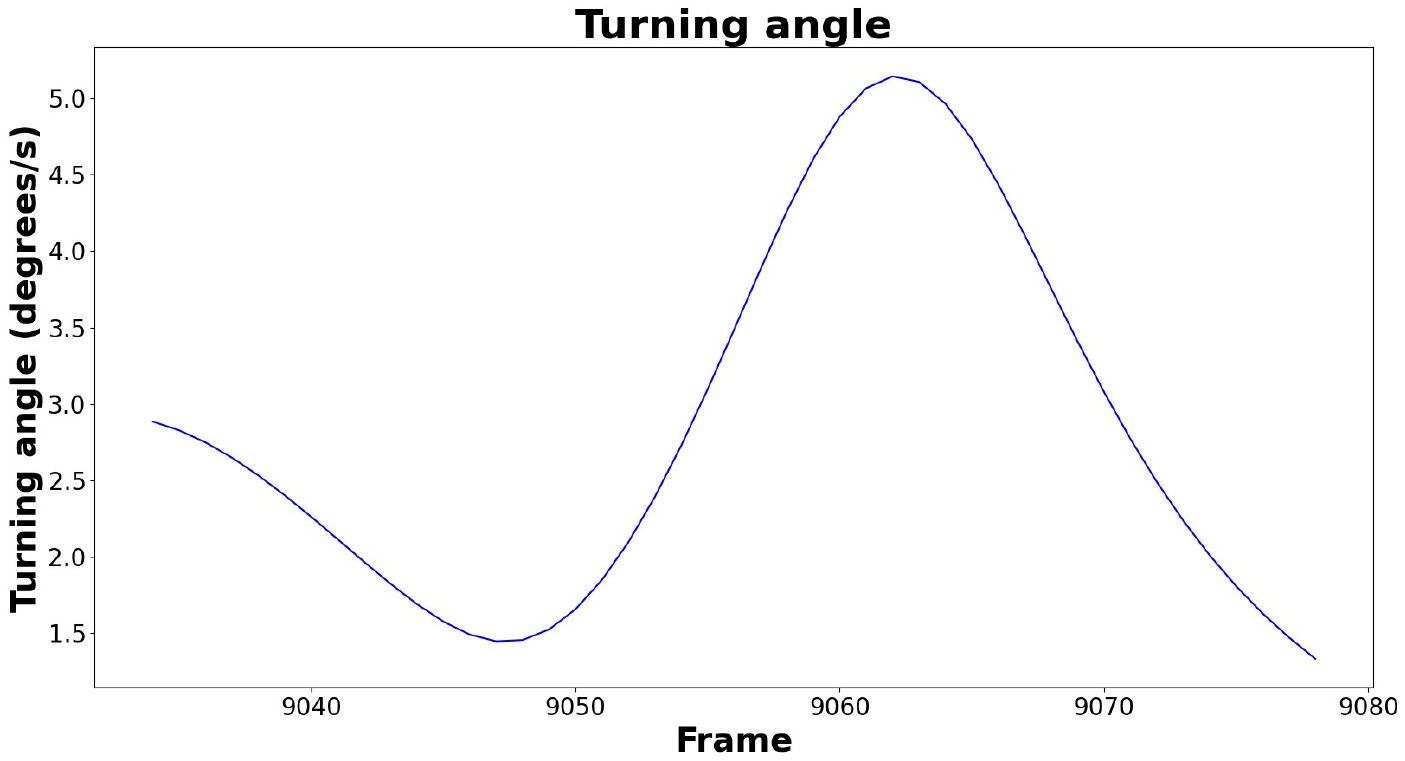}
            \caption{Turning angle}
            \label{fig:turning_angle}
        \end{minipage}
    \end{subfigure}
    
    \caption{Derived features for a single fish}
    \label{fig:derived_parameters}
\end{figure*}
 
\begin{table}[h!]
\centering

\resizebox{\columnwidth}{!}
{\begin{tblr}{
  vline{1, 3, 7} = {-}{},
  hline{2} = {3-7}{},
  hline{1,3,5,7,9} = {-}{},
  cell{1}{3,4} = {c=3}{c},
}
 \textbf{Shape} & \textbf{Color} & Initial augmented\\
  &  & Velocity [m/s] & Acceleration [$m/s^2$]  & Pitch angle [\textdegree] & Turning angle [\textdegree] \\
  
 Big Cylinder & White & 0.86 & 0.39 & 2.27 & 7.82\\
 & Yellow  & 1.33 & 0.40 & 0.76 & 9.23\\
  Cube & White & 1.07 & 0.53 & 2.62 & 10.03\\
 & Yellow  & 1.71 & 0.55 & -1.58 & 10.73\\
 Small Cylinder & White & 1.24 & 0.58 & 2.91 & 10.61\\
 & Yellow & 1.74 & 0.49 & -1.57 & 11.26

\end{tblr}}
\caption{Comparison of methods in terms of the average derived parameters velocity, acceleration, pitch angle and turning angle}

\label{table:derived_parameters}
\end{table}

An example of derived parameters for an individual fish is shown in Figure \ref{fig:derived_parameters}, describing velocity, acceleration, turning angle and pitch angle. 
The fish in question maintains a low speed through the entire trajectory, with both velocity and acceleration peaking at the end of the trajectory. 
The negative pitch angle through the entire trajectory implies that the fish approached the object but slowed down at the end of the path. 
The turning angle was consistently low. 

\subsubsection{Discussion}
The higher speed (and variability in speed) when exposed to yellow structures than for white structures could suggest more varying responses from the fish when facing this color. 
Alternatively, this could also be due to less reliable depth estimates due to a longer mean distance between the fish and the camera, affecting the derived velocity. 
The low values observed for acceleration and turning angle in particular indicate low activity levels, and could suggest that the fish did not experience stress-related responses by being in close proximity to the intrusive objects. \\

While the parameters did not present large differences in response in this particular study, the deeper insight into behaviour enabled by deriving such parameters implies that this can be a tool for capturing the variance within a fish farm over time. In addition, detecting potential outliers from expected patterns could also be useful in identifying individuals displaying abnormal patterns as these patterns may be linked to environmental concerns or welfare concerns.

\subsection{Potential method enhancements}

The proposed method has, through experimental trials, been shown to work well for estimating distances between fish and intrusive objects, as shown by comparable results in \cite{sonar_data}. 
Additionally, the method expands upon the approach in \cite{stereoYOLO} in providing the means to derive new values from both schools and individual levels. 
Several derived features, including velocity, acceleration, turning angle, and pitch angle, have been calculated and visualized, allowing for a more comprehensive view of the state of each fish and fish schools. 
In summary, this method can be used to enhance the efficiency and information used to determine the responses of fish toward intrusive objects. 

While the experiments reported here are constrained to rigid structures and no movement of the structure and, thereby the camera, the method can be useful in more dynamic operations within a fish farm. Operations frequently performed by UUVs, such as inspection and maintenance operations - including inspection of nets or mooring line - are examples where the knowledge of fish-vehicle distance and other derived parameters can be useful to ensure that the maintenance and operation methods do not negatively affect fish welfare. The method can thus provide inputs that are useful for both the design of such robots and the development of control strategies toward the development of fish-friendly autonomous operations in fish farms. 
The potential for online control from the system is underlined by the near real-time observation of nearby fish that is possible with the present method.  
Differences in the behavior of fish when exposed to potential stressors, such as low water quality, high pH values, and emerging illness, can potentially be captured by indirect behavior estimates captured by this method or similar methods, which can enable early intervention from the farmer.%is again deemed useful in this case. 

The method can still be improved by exploring new state-of-the-art approaches as more time-efficient and accurate methods become available. 
For instance, the pre-processing approaches and SuperGlue are time-consuming, and ways to decrease the time inference of these should be tested to make the method work in real-time closed loop applications. 
Further improvement could be obtained by exchanging SuperGlue, which may cause less reliable results since it does not estimate feature correspondences for all detected tails.
A possible better alternative could be a feature correspondence deep learning estimator better suited for underwater scenarios.\\
%SuperGlue also does not estimate feature correspondences for all detected tails, which may cause less reliable results as fewer fish are then being tracked. 

If the method could be rendered less resource demanding and computationally intensive, it could be adapted to run on more accessible hardware platforms, such as Raspberry Pi, Nvidia Jetson or similar, making it accessible for both control and real-time monitoring in industrial fish farms.
Finally, by combining this approach with other sensors, such as the one in \cite{sonar_data} , fish behavior monitoring can be made more accurate and precise, simpler, more efficient, and user-friendly, and thereby indirectly provide better welfare for the future fish farming industry. 
%In the future, this method could be implemented on more accessible hardware platforms, such as Raspberry Pi, Nvidia Jetson or similar, making it accessible for both control and real-time monitoring in industrial fish farms.

\section{Conclusions and Future work}
\label{sec:conc}
In this study, a complete object detection, tracking, and 3D position estimation pipeline specifically designed for tracking the caudal fins of salmon in industrial sea cages has been proposed. The method has been successfully applied to relevant data from an industrial fish cage and has provided interesting results regarding differences in fish behavior when exposed to varying intrusive objects. Fish were exposed to stationary objects with different shapes, sizes, and colors, and their response varied based on size and color. The fish stayed further away from yellow and larger structures than white and smaller ones. 
Results concerning the structure's shape were inconclusive. 
The fish did not seem to react through increased velocity, acceleration, pitch angle, or turning angle, possibly indicating that stationary objects do not feel like a significant threat to salmon. 

For future research, applying the method to other scenarios with dynamic objects that better mimic the operations in industrial fish farms, and considering the speed, acceleration, pitch angle, and turning angle when compared to such still-standing objects may provide additional insight into fish-vehicle dynamics. 
Moreover, the approach of combining several types of sensors (e.g., stereo cameras and sonar) in a common package and using sensor fusion to combine their data could prove a more robust method for observing fish responses to intrusive objects. 
Such tools, when put into action and tested across reasonable scenarios, can provide important knowledge for developing the aquaculture practices of the future. 

\section*{Acknowledgment}
The authors would like to thank Qin Zhang, Linn Evjemo, Damiano Varagnolo for their valuable inputs and discussions on the method implementation and data processing.

\FloatBarrier

\bibliographystyle{IEEEtran}
\bibliography{./IEEEabrv,references}
%%%%%%%%%%%%%%%%%%%%%%%%%%%%%%%%%%%%%%%%%%%%%%%%%%%%%%%%%%%%%%%%%%%%%%%%%%%%%%%%
\appendices

\section{Distance Estimates and FPS}
\FloatBarrier

\begin{table}[h!]
\resizebox{\columnwidth}{!}
{\begin{tblr}{
  vline{1, 3, 7} = {-}{},
  hline{2} = {3-6}{},
  hline{1,3,5,7,9} = {-}{},
  cell{1}{3,4} = {c=3}{c},
}
 \textbf{Shape} & \textbf{Color} & \textbf{Average minimum distance [m]} & &\\ 
 
  &  & Initial & {MO-WT} & {Initial \\ augmented}  & {MO-WT \\augmented} \\
 Big Cylinder & White & 1.18 & 1.16 & 1.18 & 1.18\\
 & Yellow  &  1.66 & 1.53 & 1.74 & 1.74 \\
  Cube & White  & 1.08 & 1.04 & 1.11 & 1.11\\
 & Yellow & 1.43 & 1.41 & 1.42 & 1.43 \\
 Small Cylinder & White  & 1.12 & 1.11 & 1.07 & 1.08\\
 & Yellow  &  1.40 & 1.39 & 1.44 & 1.44

\end{tblr}}
\caption{Comparison of methods in terms of average minimum distance}

\label{table:comparison_models_min_dist}
\end{table}

\begin{table}[h!]
\resizebox{\columnwidth}{!}
{\begin{tblr}{
  vline{1, 3, 7} = {-}{},
  hline{2} = {3-6}{},
  hline{1,3,5,7,9} = {-}{},
  cell{1}{3,4} = {c=3}{c},
}
 \textbf{Shape} &\textbf{Color} & \textbf{Aggregated maximum distance [m]}\\
  &  & Initial & {MO-WT} & {Initial \\ augmented} & {MO-WT \\ augmented} \\
  
 Big Cylinder & White & 2.68 & 2.74 & 2.70 & 2.71\\
 & Yellow & 2.67 & 2.63 & 2.66 &  2.67\\
  Cube & White  & 2.43 & 2.45 & 2.50 & 2.50\\
 & Yellow & 2.63 & 2.65 & 2.67 & 2.67 \\
 Small Cylinder & White  & 2.58 & 2.56 & 2.60 & 2.60\\
 & Yellow  &  2.41 & 2.43 & 2.55 & 2.55

\end{tblr}}
\caption{Comparison of methods in terms of average aggregated maximum distance}
\end{table}

\begin{table}[h!]
\resizebox{\columnwidth}{!}
{
\begin{tblr}{
    vline{2,3} = {1, 3-5, 7-9}{},
    hline{1,2, 3, 4,5, 6,7,8, 9, 10, 11, 12, 13, 14, 15} = {-}{},
    vline{1, 4} = {-}{}}

    \textbf{Method} & \textbf{Image width} & \textbf{FPS} \\[1.3ex]
    \textbf{Foreground caudal fins dataset}\\[1.3ex]
     Initial    & 640    & 13.1 \\[1.3ex]
    MO-WT augmented & 1024 & 2.7 \\[1.3ex]
    MO-WT augmented + RAFT-Stereo & 1024 & 1.8 \\[1.3ex] 
    \textbf{Full caudal fins dataset}\\[1.3ex]
    Initial augmented & 1024 & 12.8 \\[1.3ex]
    MO-WT augmented &  1024 &  2.7 \\[1.3ex]

    \end{tblr}}
    \caption{Average FPS for different object detection approaches}

\label{table:fps}
\end{table}

\FloatBarrier
\section{Detected and Tracked Fish}

% \begin{table}[h!]
% \centering
% \resizebox{\columnwidth}{!}{
% \begin{tblr}{
%   vline{1, 3, 8} = {-}{},
%   hline{2} = {3-7}{},
%   hline{1,3,5,7,9} = {-}{},
%   cell{1}{3,4,5,6,7} = {c=3}{c},
% }
% \textbf{Shape} & \textbf{Color} & \textbf{Average number of frames without tracked fish} & & & &\\
%  & & \textbf{Initial} & \textbf{MO-WT} & \textbf{Initial augmented} & \textbf{MO-WT augmented} \\
% Big Cylinder & White & 83 & 86 & 68 & 66\\
%  & Yellow & 1439 & 1583 & 910 & 901\\
% Cube & White & 129 & 76 & 76 & 72\\
%  & Yellow & 1123 & 925 & 621 & 619 \\
% Small Cylinder & White & 414 & 297 & 247 & 276\\
%  & Yellow & 3712 & 3471 & 2503 & 2551 \\
% \end{tblr}}
% \caption{Comparison of methods in terms of the average number of frames without tracked fish}
% \label{table:comparison_models_untracked_frames}
% \end{table}

%test

\begin{table}[h!]
\centering
\resizebox{\columnwidth}{!}{
\begin{tblr}{
  vline{1,3,7} = {-}{},
  hline{1,3,5,7,9} = {-}{},
  hline{2} = {3-6}{},
}
\textbf{Shape} & \textbf{Color} &
\SetCell[c=4]{c}{\textbf{Average number of frames without tracked fish}} \\

& &
\textbf{Initial} &
\textbf{MO-WT} &
\textbf{Initial augmented} &
\textbf{MO-WT augmented} \\

Big Cylinder & White & 83 & 86 & 68 & 66 \\
& Yellow & 1439 & 1583 & 910 & 901 \\

Cube & White & 129 & 76 & 76 & 72 \\
& Yellow & 1123 & 925 & 621 & 619 \\

Small Cylinder & White & 414 & 297 & 247 & 276 \\
& Yellow & 3712 & 3471 & 2503 & 2551 \\
\end{tblr}
}
\caption{Comparison of methods in terms of the average number of frames without tracked fish}
\label{table:comparison_models_untracked_frames}
\end{table}

\begin{table}[h!]
\centering

\resizebox{\columnwidth}{!}
{\begin{tblr}{
  vline{1, 3, 7} = {-}{},
  hline{2} = {3-6}{},
  hline{1,3,5,7,9} = {-}{},
  cell{1}{3,4} = {c=3}{c},
}
 \textbf{Shape} & \textbf{Color} & \textbf{Detected fish/Tracked fish} \\
  &  & Initial & {MO-WT} & {Initial \\ augmented}  & {MO-WT \\ augmented} \\
 Big Cylinder & White & 7476/4058  & 8784/4700 & 19834/4913 & 19243/4942\\
 & Yellow  &  7840/3042 & 10166/3140 & 18049/4040 & 19328/4123\\
  Cube & White  & 9761/4646 & 11928/4635 & 19894/6100 & 18056/6034\\
 & Yellow  & 7430/2893 & 9326/3039 & 20232/4286 & 17690/4242 \\
 Small Cylinder & White  & 7023/3393 & 7145/2918 & 16086/4489 & 14655/4340\\
 & Yellow  &  3679/1413 & 4588/1444 & 12886/2291 & 11920/2219
\end{tblr}}
\caption{Comparison of methods in terms of the average number of detected and tracked fish}
\label{table:comparison_models_num_fish}
\end{table}

\FloatBarrier
\newpage
\section{Derived Parameters}
\begin{table}[h!]

\resizebox{\columnwidth}{!}
{\begin{tblr}{
  vline{1, 7, 3} = {-}{},
  hline{2} = {3-6}{},
  hline{1,3,5,7,9} = {-}{},
  cell{1}{3,4} = {c=3}{c},
}
 \textbf{Shape} & \textbf{Color} & \textbf{Average velocity [m/s]}\\
  &  & Initial & {MO-WT} & {Initial \\ augmented}  & {MO-WT \\ augmented} \\
  
 Big Cylinder & White & 0.68 & 0.80 & 0.86 & 0.87\\
 & Yellow  & 1.33 & 1.31 & 1.33 & 1.34\\
  Cube & White  & 1.16 & 1.00 & 1.07 & 1.06\\
 & Yellow  & 1.78 & 1.69 & 1.71 & 1.70 \\
 Small Cylinder & White  & 1.16 & 1.17 & 1.24 &1.24\\
 & Yellow  &  1.68 & 1.60 & 1.74 & 1.71
\end{tblr}}
\caption{Comparison of methods in terms of average velocity}
\label{table:comparison_models_vel}
\end{table}

\begin{table}[h!]
\resizebox{\columnwidth}{!}
{\begin{tblr}{
  vline{1, 3, 7} = {-}{},
  hline{2} = {3-6}{},
  hline{1,3,5,7,9} = {-}{},
  cell{1}{3,4} = {c=3}{c},
}
 \textbf{Shape} & \textbf{Color} & \textbf{Average acceleration [$m/s^2$]}\\
  &  & Initial & {MO-WT} & {Initial \\ augmented}  & {MO-WT \\ augmented} \\
  
 Big Cylinder & White & 0.16 & 0.29 & 0.39 & 0.29\\
 & Yellow  & 0.29 & 0.13 & 0.40 & 0.36\\
  Cube & White  & 0.53 & 0.51 & 0.53 & 0.52\\
 & Yellow  & 0.30 & 0.30 & 0.55 & 0.59\\
 Small Cylinder & White  & 0.22 & 0.38 & 0.58 & 0.64\\
 & Yellow  &  0.28 & 0.34 & 0.49 & 0.28
\end{tblr}}
\caption{Comparison of methods in terms of average acceleration}
\label{table:comparison_models_acc}
\end{table}

\begin{table}[h!]

\resizebox{\columnwidth}{!}
{
\begin{tblr}{
  vline{1, 3, 7} = {-}{},
  hline{2} = {3-6}{},
  hline{1,3,5,7,9} = {-}{},
  cell{1}{3,4} = {c=3}{c},
}
 \textbf{Shape} & \textbf{Color} & \textbf{Average pitch angle [\textdegree]}\\
  &  & Initial & {MO-WT} & {Initial \\ augmented}  & {MO-WT \\ augmented} \\
  
 Big Cylinder & White & 2.83 & 3.34 & 2.27 & 2.63\\
 & Yellow  & 1.08 & -1.63 & 0.76 & 0.90\\
  Cube & White  & 2.94 & 2.73 & 2.62 & 2.84\\
 & Yellow  & -1.75 & -2.03 & -1.58 & -1.72\\
 Small Cylinder & White  & 4.15 & 4.94 & 2.91 & 3.35\\
 & Yellow  & -1.64 & -1.45 & -1.57 & -1.87

\end{tblr}}
\caption{Comparison of methods in terms of average pitch angle}
\label{table:comparison_models_pitch}
\end{table}
\newpage

\begin{table}[ht!]
\resizebox{\columnwidth}{!}
{
\begin{tblr}{
  vline{1, 3, 7} = {-}{},
  hline{2} = {3-6}{},
  hline{1,3,5,7,9} = {-}{},
  cell{1}{3,4} = {c=3}{c},
}
 \textbf{Shape} & \textbf{Color} & \textbf{ Average turning angle [\textdegree]} \\
  &  & Initial & {MO-WT} & {Initial \\ augmented}  & {MO-WT \\ augmented} \\
  
 Big Cylinder & White & 5.48 & 6.07 & 7.82 & 7.83\\
 & Yellow  & 8.45 & 8.54 & 9.23 & 9.20\\
  Cube & White  & 10.23 & 9.44 & 10.03 & 9.93\\
 & Yellow  & 10.60 & 9.96 & 10.73 & 10.60\\
 Small Cylinder & White  & 9.80 & 9.02 & 10.61 & 10.55\\
 & Yellow  &  10.36 & 9.84 & 11.26 & 10.97

\end{tblr}}
\caption{Comparison of methods in terms of average turning angle}
\label{table:comparison_models_turn}
\end{table}
\end{document}